\documentclass[aps,prd,preprint,nofootinbib,showpacs]{revtex4-1}
\usepackage{graphicx}
\usepackage{epsfig}
\usepackage{amssymb}
\usepackage{color}
\usepackage{epstopdf}
\usepackage{dcolumn}
\usepackage{amsmath}
\usepackage{bm}
\usepackage{dcolumn}
\usepackage{txfonts}
\begin{document}

\title{Strong coupling constant and heavy quark masses in 
(2+1)-flavor QCD}

\author{
P. Petreczky$^{\rm a}$, 
J. H. Weber$^{\rm b,c}$
}
\affiliation{
$^{\rm a}$ Physics Department, Brookhaven National Laboratory, Upton, NY 11973, USA \\
$^{\rm b}$ Department of Computational Mathematics, Science and Engineering, and 
Department of Physics and Astronomy,
Michigan State University, East Lansing, MI 48824, USA\\
$^c$ Exzellenzcluster Universe, Technische Universit\"{a}t M\"{u}nchen, D-85748 Garching, Germany
}

\begin{abstract}
We present a determination of the strong coupling constant and heavy quark masses
in (2+1)-flavor QCD using lattice calculations of the moments of the pseudo-scalar 
quarkonium correlators at several values of the heavy valence quark mass with Highly Improved Staggered Quark (HISQ) action.
We determine the strong coupling constant in the $\overline{MS}$ scheme at four low-energy scales 
corresponding to $m_c$, $1.5m_c$, $2m_c$, and $3m_c$, with $m_c$ being the charm
quark mass. The novel feature of our analysis that up to eleven lattice spacings are used
in the continuum extrapolations, with the smallest lattice spacing being $0.025$ fm.
We obtain $\Lambda_{\overline{MS}}^{n_f=3}=298 \pm 16$ MeV, which
is equivalent to $\alpha_s(\mu=M_Z,n_f=5)=0.1159(12)$.
For the charm and bottom quark masses in the $\overline{MS}$ scheme, we obtain: 
$m_c(\mu=m_c,n_f=4)=1.265(10)$ GeV and 
$m_b(\mu=m_b,n_f=5)=4.188(37)$ GeV.
\end{abstract}
\date{\today}

\maketitle

\section{Introduction}
In recent years there has been an extensive effort toward
the accurate determination of QCD parameters since the
precise knowledge of these parameters is important for testing the predictions of the Standard Model.
Two important examples are the sensitivity of the Higgs branching ratios to the heavy quark masses
and the strong coupling constant \cite{Dawson:2013bba,Lepage:2014fla} and the stability of the Standard Model
vacuum \cite{Buttazzo:2013uya,Espinosa:2013lma}.
Lattice calculations play an increasingly important role in the determination of the QCD parameters
as these calculations become more and more precise with the advances in computational approaches.

The strong coupling constant $\alpha_s$ has been known for a long time.
The current Particle Data Group (PDG) value $\alpha_s(M_Z)=0.1181(11)$ \cite{PDG18} has small
errors  suggesting that the uncertainties are well under control.
However, a closer inspection of the PDG averaging procedure shows
that the individual $\alpha_s$ determinations have rather large errors.
The PDG's $\alpha_s$ determinations are grouped into different
categories according to the observables used in the analysis \cite{PDG18}.
Individual determinations within each category often have very different errors
and different central values suggesting that not all the sources of
errors are fully understood. In particular, there are determinations of $\alpha_s$
from jet observables \cite{Abbate:2010xh,Hoang:2015hka} that are in clear tension with the PDG average.

The new lattice average of $\alpha_s$ provided by the Flavor Lattice Averaging Group (FLAG) 
agrees well with the PDG
average \cite{FLAG19}. The $\alpha_s$ determinations that enter the FLAG average agree well
with each other \cite{FLAG19}, though the $\alpha_s$ value obtained from the static quark anti-quark
energy is lower compared to the other determinations. 
A noticeable difference of the new FLAG averaging procedure compared to the previous one \cite{FLAG16} is the decreased weight
of the $\alpha_s$ determination from the moments of quarkonium correlators, due to the more conservative
error assigned by FLAG.
Therefore, improved
calculations of $\alpha_s$ from the moments of quarkonium correlators are desirable.

The running of the strong coupling constant at lower energy scales is also interesting. 
For example, for testing the weak coupling approach to QCD thermodynamics
through comparison to lattice QCD results 
\cite{Bazavov:2018wmo,Bazavov:2017dsy,Bazavov:2016uvm,Berwein:2015ayt,Ding:2015fca,Haque:2014rua,Bazavov:2013uja}
one needs to know the coupling
constant at a relatively low-energy scale of approximately $\pi T$, with $T$ being the temperature.
The analysis of the $\tau$ decay offers the possibility
of $\alpha_s$ extraction at a low-energy scale, but there are large systematic
uncertainties due to different ways of organizing the perturbative expansion
in this method (see Refs.~\cite{Pich:2016bdg,Boito:2016oam,Boito:2016hzq} and
references therein for recent work on this topic). Lattice QCD calculations, on the other hand,
are well suited to map out the running of $\alpha_s$ at low energies.

Lattice determination of the charm quark mass has significantly improved over the years.
The current status of charm quark mass determination on the lattice is reviewed in
the new FLAG report \cite{FLAG19}. Though the current FLAG average for the charm quark mass
has smaller errors than the PDG value, there are some inconsistencies
in different lattice determinations (cf. Fig. 5 in Ref. \cite{FLAG19}). Therefore, additional lattice
calculations of the charm quark mass could be useful. 

Determination of the bottom quark mass is difficult in 
the lattice simulations due to the large
discretization errors caused by powers of $m_h a$, where $m_h$ is the bare mass of the heavy quarks.
One needs a small lattice spacing to control the corresponding discretization
errors. One possibility to deal with this problem is to perform
calculations with heavy quark masses smaller than the bottom quark mass 
and extrapolate to the bottom quark mass guided by the heavy quark effective theory 
\cite{McNeile:2010ji,Bazavov:2018omf}.
The current status of the bottom quark mass calculations on the lattice 
is also reviewed in the new FLAG report \cite{FLAG19}. It is clear that 
the bottom quark mass determination will benefit from new calculations
on finer lattices.

In this paper we report on the calculations of $\alpha_s$ and the heavy quark masses 
in (2+1)-flavor QCD using the Highly Improved Staggered
Quark (HISQ) action and moments of pseudo-scalar quarkonium correlators. We  extend
the previous work reported in Ref. \cite{Maezawa:2016vgv} by considering several valence heavy quark masses,
drastically reducing the statistical errors and most notably by extending the lattice calculations 
to much finer lattices. 
This paper is organized as follows. In Section \ref{sec:lat_setup} we introduce the details of the lattice setup.
In Section \ref{sec:mom} we discuss the moments of quarkonium correlators and our main numerical results.
The extracted values of the strong coupling constant and the heavy quark masses are discussed in Section \ref{sec:asmh}
and compared to other lattice and non-lattice determinations.
The paper is concluded in Section~\ref{sec:conclusion}. Some technical details of the calculations are given in the Appendices.

\section{Lattice setup and details of analysis}
\label{sec:lat_setup}
To determine the heavy quark masses and the strong coupling constant,
we calculate the pseudo-scalar quarkonium correlators in 
(2+1)-flavor lattice QCD.
As in our previous study, we take advantage of the gauge configurations 
generated using the tree-level improved
gauge action \cite{Luscher:1984xn} and the Highly Improved Staggered Quark (HISQ) action \cite{Follana:2006rc}
by the HotQCD collaboration \cite{Bazavov:2014pvz}. The strange quark mass, $m_s$, was fixed to its
physical value, while for the light (u and d) quark masses the value $m_l=m_s/20$ was used.
The latter corresponds to the pion mass  $m_{\pi}=161$ MeV in the continuum limit, i.e. the sea quark
masses are very close to the physical value. This is the same set of gauge configurations as used
in Ref. \cite{Maezawa:2016vgv}.
We use additional HISQ gauge configurations with
light sea quark masses $m_l=m_s/5$, \mbox{i.e.}, the pion mass of $322$ MeV, at five lattice spacings corresponding
to the lattice gauge couplings $\beta=10/g_0^2=7.03,~7.825,~8.0,~8.2$, and $8.4$, 
generated for the study of the QCD equation of state at high temperatures \cite{Bazavov:2017dsy}.
This allows us to perform calculations at three smaller lattice spacings, namely, $a=0.035$, $a=0.03$, and $0.025$ fm
and also check for sensitivity of the results to the light sea quark masses. 
As we will see later, the larger than the physical sea quark mass has no effect on the moments
of quarkonium correlators. 

For the valence heavy quarks we use the HISQ action with the so-called
$\epsilon$-term \cite{Follana:2006rc}, which removes the tree-level discretization effects
due to the large quark mass up to ${\cal O}((a m)^4)$. The HISQ action with the $\epsilon$-term
turned out to be very effective for treating the charm quark on the lattice
\cite{Follana:2006rc,Follana:2007uv,Davies:2010ip,Bazavov:2014wgs,Bazavov:2014cta}.
The lattice spacing in our calculations has been fixed using the $r_1$ scale
defined in terms of the energy of a static quark anti-quark pair $V(r)$ as
\begin{equation}
\left. r^2 \frac{d V}{d r} \right|_{r=r_1}=1.0.
\end{equation}
We use the value of $r_1$ determined in Ref.~\cite{Bazavov:2010hj}
using the pion decay constant as an input:
\begin{equation}
r_1=0.3106 \, (18) \ {\rm fm}.
\label{r1}
\end{equation}
In the above equation all the sources of errors in Ref.~\cite{Bazavov:2010hj} have been
added in quadrature. 
The above value of $r_1$ corresponds to the value of the scale parameter 
determined from the Wilson flow $w_0=0.1749(14)$ fm \cite{Bazavov:2014pvz}. 
This agrees very well with the determination of the Wilson flow parameter 
by the BMW collaboration $w_0=0.1755(18)(4)$ fm \cite{Borsanyi:2012zs}.
It is also consistent with the value $r_1=0.3133(23)(3)$ fm reported by the HPQCD collaboration within errors \cite{Davies:2009tsa}. 
It turns out that the value of $r_1/a$ does not change within errors when increasing the sea quark mass from $m_s/20$
to $m_s/5$ \cite{Bazavov:2017dsy}, and therefore, the lattice results on $r_1/a$ at the two quark masses can be combined
to obtain the parametrization of $r_1/a$ as function of $\beta$ \cite{Bazavov:2017dsy}. We use this parameterization
to determine the lattice spacing.

\begin{table}
\begin{tabular}{ccccccc}
\hline\hline
\(\beta\) & \(\frac{m_\ell}{m_s}\)&~~ lattice~~ &~~ $a^{-1}$ GeV~~~ &~ $L_s$ fm~~  & $a m_{c0}$ & $a m_{b0}$ \\
\hline
6.740 & 0.05 & $48^4$           &   1.81 & 5.2 & 0.5633(10)  &            \\
6.880 & 0.05 & $48^4$           &   2.07 & 4.6 & 0.4800(10)  &            \\
7.030 & 0.05 & $48^4$           &   2.39 & 4.0 & 0.4047(9)   &            \\
7.150 & 0.05 & $48^3 \times 64$ &   2.67 & 3.5 & 0.3547(9)   &            \\
7.280 & 0.05 & $48^3 \times 64$ &   3.01 & 3.1 & 0.3086(13)  &            \\
7.373 & 0.05 & $48^3 \times 64$ &   3.28 & 2.9 & 0.2793(5)   &            \\
7.596 & 0.05 & $64^4$           &   4.00 & 3.2 & 0.2220(2)   & 1.019(8)   \\
7.825 & 0.05 & $64^4$           &   4.89 & 2.6 & 0.1775(3)   & 0.7985(5)  \\
\hline
7.030 & 0.20 & $48^4$           &   2.39 & 4.0 & 0.4047(9)   &            \\
7.825 & 0.20 & $64^4$           &   4.89 & 2.6 & 0.1775(3)   & 0.7985(5)  \\
8.000 & 0.20 & $64^4$           &   5.58 & 2.3 & 0.1495(6)   & 0.6710(6)  \\
8.200 & 0.20 & $64^4$           &   6.62 & 1.9 & 0.1227(3)   & 0.5519(6)  \\
8.400 & 0.20 & $64^4$           &   7.85 & 1.6 & 0.1019(27)  & 0.4578(6)  \\
\hline
\end{tabular}
\caption{The lattice gauge couplings, the light quark masses, the lattice sizes, the inverse lattice spacings, the spatial lattice size in fm, 
the bare charm quark
mass, and the bare bottom quark mass used in our calculations. The upper part of the table corresponds to the lattices
from Ref. \cite{Bazavov:2014pvz}, while the lower part corresponds to lattices from Ref. \cite{Bazavov:2017dsy}.}
\label{tab:latpar}
\end{table}

We calculate pseudo-scalar meson correlators for different heavy quark masses
using random color wall sources \cite{Chakraborty:2014aca}. This reduces
the statistical errors in our analysis by an order of magnitude compared to the previous
study \cite{Maezawa:2016vgv}. 
Here, we consider several quark masses in the region 
between the charm and bottom quark, namely, $m_h=m_c,~1.5m_c,~2m_c,~3m_c,~4m_c$, and $m_b$.
This helps us to study the running of $\alpha_s$ at
low energies and provides additional cross-checks on the error analysis.
The bare quark mass $m_{c0}$ that corresponds to the physical charm quark mass has been
determined in Ref. \cite{Maezawa:2016vgv} by fixing the spin-average 1S charmonium mass,
$(3 M_{J/\psi}+M_{\eta_c})/4$, to its physical value. For $\beta \le 7.825$ we use the same values of $m_{c0}$ in this work.
For the three largest values of $\beta$ we determined the bare charm quark mass
by requiring that the mass of $\eta_c$ meson obtained on the lattice in physical units agrees with the corresponding PDG value.
This is equivalent to fixing $m_{c0}$ by the spin averaged 1S charmonium mass since the hyperfine splitting 
$M_{J/\psi}-M_{\eta_c}$ is expected to be well reproduced on these fine lattices.
We determine the $b$ quark mass at each value of gauge coupling $\beta$
by performing calculations at several values
of the heavy quark mass near the $b$ quark mass and linearly interpolating to find
the quark mass at which the pseudo-scalar mass is equal to the physical mass of $\eta_b$ meson from PDG.
In these calculations we use both
random color wall sources and corner wall sources. 
In Table \ref{tab:latpar} we summarize the gauge ensembles used in our study, %the number of measurements for each $\beta$ 
as well as the values of the bare charm and bottom quark masses.

\begin{figure}
\includegraphics[width=10cm]{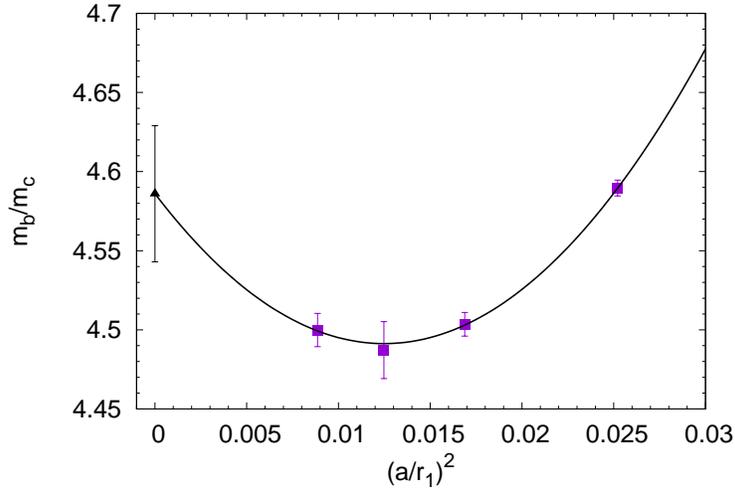}
\caption{The ratio $m_b/m_c$ as function of $(a/r_1)^2$ together with
the corresponding continuum value. The solid line is the fit to the $a^2+a^4$ form.
}
\label{fig:mb}
\end{figure}

In Fig. \ref{fig:mb} we show
our results for the ratio $m_b/m_c$ as a function of the lattice spacing. 
The ratio does not show a simple scaling with $a^2$, and therefore we fit
the data with the $a^2+ a^4$ form, which leads to a continuum value of
\begin{equation}
m_b/m_c=4.586(43)
\label{mb_by_mc}
\end{equation}
The above result agrees with the previous (2+1)-flavor HISQ analysis that was based on
extrapolations to the bottom quark mass region and resulted in $m_b/m_c=4.528(57)$ \cite{Maezawa:2016vgv}.
It also agrees with the HPQCD result $m_b/m_c=4.528(54)$ \cite{Chakraborty:2014aca} as well as with the new
Fermilab-MILC-TUMQCD result $m_b/m_c=4.577(8)$ \cite{Bazavov:2018omf} both obtained in 2+1+1 flavor QCD.

\section{Moments of quarkonium correlators and QCD parameters}
\label{sec:mom}
We consider moments of the pseudo-scalar quarkonium correlator,
which are defined as
\begin{equation}
G_n=\sum_t t^n G(t),~G(t)=a^6 \sum_{\mathbf{x}} (a m_{h0})^2 \langle j_5(\mathbf{x},t) j_5(0,0) \rangle.
\end{equation}
Here $j_5=\bar \psi \gamma_5 \psi $ is the pseudo-scalar current and $m_{h0}$ is the lattice heavy quark mass.
To take into account the periodicity of the lattice of temporal size $N_t$ the above definition of the moments can
be generalized as follows:
\begin{equation}
G_n=\sum_t t^n (G(t)+G(N_t-t)).
\label{eq:latmom}
\end{equation}
The moments $G_n$ are finite only for $n \ge 4$ ($n$ even), since the correlation function 
diverges as $t^{-4}$ for small $t$. Furthermore, the moments $G_n$ do not
need renormalization because
 the explicit factors of the quark mass are included in
their definition \cite{Allison:2008xk}.
They can be calculated in perturbation theory in the $\overline{MS}$ scheme
\begin{equation}
G_n=\frac{g_n(\alpha_s(\mu),\mu/m_h)}{a m_h^{n-4}(\mu_m)}.
\end{equation}
Here, $\mu$ is the $\overline{MS}$ renormalization scale. The scale, $\mu_m$, at which the $\overline{MS}$ 
heavy quark mass is defined can be different from $\mu$ \cite{Dehnadi:2015fra}, though most studies
assume $\mu_m=\mu$.
The coefficient $g_n(\alpha_s(\mu),\mu/m_h)$ is calculated up to 4-loop, i.e., up to order $\alpha_s^3$
\cite{Sturm:2008eb,Kiyo:2009gb,Maier:2009fz}.
Given the lattice data on $G_n$ one can extract $\alpha_s(\mu)$ and $m_c(\mu)$ from the above
equation. However, as was pointed out in Ref.~\cite{Allison:2008xk} it is more practical to
consider the reduced moments
\begin{equation}
R_n =\left\{ \begin{array}{ll}
G_n/G_n^{(0)} & (n=4) \\
\left(G_n/G_n^{(0)}\right)^{1/(n-4)} & (n\ge6) \\
\end{array} \right.
\label{eq:redmom},
\end{equation}
where $G_n^{(0)}$ is the moment calculated from the free correlation function.
The lattice artifacts largely cancel out in these reduced moments.
Our numerical results on $R_n$ and some of the relevant ratios,
e.g., $R_8/R_{10}$, $R_8/m_{c0}$, etc., are given in Appendix~\ref{app:A}.
From the tables one can clearly see that the statistical errors
are tiny and can be neglected. 
The light sea quark masses $m_l$ used in our calculations on the three finest
lattices are about five times larger than the physical ones, and the effect of this
on the reduced moments needs to be investigated. At two values of the gauge
coupling, namely, $\beta=7.03$, and $7.825$, we have calculated the reduced moments
for two values of the light sea quark mass, $m_l=m_s/20$, and $m_l=m_s/5$. From
the corresponding results on the reduced moments given in Appendix ~\ref{app:A},
we see that the effect of the light sea quark mass
is of the order of statistical errors and therefore will
be neglected in the analysis.

The dominant errors
in our calculations are the errors due to finite volume and
the errors induced by mistuning of the heavy quark mass. 
To estimate the finite volume effects we use the moments of the correlators
estimated in the free theory on the lattices that are used in our simulations as well as in the infinite volume limit. 
The difference between
the two results, $\delta_V G_n^{0}=G_n^{0,V}-G_n^{0}$
could be used as an estimate of the finite volume errors in $G_n$.
If the finite volume errors were the same in the free theory and in the
interacting theory, the reduced moments $R_n$  would have no finite volume errors (the finite volume errors would cancel between
the numerator and denominator). In reality, the finite volume effects are different in the free theory and in the interacting
theory, and $R_n$ will be affected by the finite volume.
We assume that the finite
volume effects are similar in size to those in the free theory but could be different in the absolute value and in the sign, and thus
would not cancel in the reduced moments, i.e., it is assumed that the finite volume errors in $R_n$ are given by 
$\delta_V G_n^0/G_n^0$. It is known that the finite size effects
in the interacting theory are much smaller \cite{McNeile:2010ji,Chakraborty:2014aca} than in the free theory,
so the above estimate of the finite volume effects is rather conservative.
The finite volume errors at different lattice spacings are correlated, since they are expected
to be a smooth function of $\beta$. However, since the physical volume is deceasing in our calculations
with the decreasing lattice spacing, the errors are not $100\%$ correlated. Since 
we only have a rough estimate of the finite volume errors a proper evaluation of the correlations
is not possible. Therefore, in our analysis we assumed that finite volume errors are uncorrelated. 
We also considered the possibility that all finite volume errors are $100\%$ correlated. As expected
that resulted in a larger error for a given continuum extrapolation. However, the corresponding
increase in the errors was still considerably smaller than the finite error estimate of the continuum
result that combines many fits. Therefore, we concluded that at present it is justified to treat
the finite volume errors as uncorrelated.

To estimate the errors due to the mistuning of the heavy quark mass, we performed
interpolation of $R_n$ in the heavy quark mass and examined the changes in $R_n$
when the value of the heavy quark mass was varied by one sigma.
The systematic errors due to the finite volume and mistuning of the heavy quark mass are summarized in Appendix~\ref{app:A}.

It is straightforward to write down the perturbative expansion for $R_n$:
\begin{eqnarray}
R_n &=& \left\{ \begin{array}{ll}
r_4 & (n=4) \\
r_n \cdot \left({m_{h0}}/{m_h(\mu)}\right) & (n\ge6)\\
\end{array}\right. , \label{rn_pert}\\
r_n &=& 1 + \sum_{j=1}^3 r_{nj}(\mu/m_h) \left(\frac{\alpha_s(\mu)}{\pi}\right)^j.
\end{eqnarray}
From the above equations, it is clear that $R_4$ as well as the ratios $R_6/R_8$ and $R_8/R_{10}$
are suitable for the extraction of the strong coupling constant $\alpha_s(\mu)$, while
the ratios $R_n/m_{h0}$ with $n\ge 6$ are suitable for extracting the heavy quark mass $m_h(\mu)$.
In our analysis, we choose the renormalization scale $\mu=m_h(m_h)$. With this choice, the expansion coefficients,
$r_{nj}(\mu/m_h)$,
are just simple numbers that are given in Table \ref{tab:rn}.
This choice of the renormalization scale has the advantage that the expansion coefficients
are never large. If the renormalization scale is different from $m_h$, the scale dependence
of $m_h$ needs to be taken into account, which increases the uncertainty of the perturbative
result \cite{Dehnadi:2015fra}.
\begin{table}
\begin{tabular}{|c|lll|}
\hline
n & $r_{n1}$ & $r_{n2}$ &  $r_{n3}$  \\
\hline
4  &  2.3333 & -0.5690  &  1.8325    \\
6  &  1.9352 &  4.7048  & -1.6350    \\
8  &  0.9940 &  3.4012  &  1.9655    \\
10 &  0.5847 &  2.6607  &  3.8387    \\
\hline
\end{tabular}
\caption{The coefficients of the perturbative expansion of $R_n$.}
\label{tab:rn}
\end{table}
There is also a non-perturbative contribution to the moments proportional
to the gluon condensate \cite{Broadhurst:1994qj}. We included this contribution 
at tree level using
the value 
\begin{equation}
\langle \frac{\alpha_s}{\pi} G^2 \rangle = 0.006 \pm 0.012
\label{cond}
\end{equation}
from the analysis of $\tau$ decay
\cite{Geshkenbein:2001mn}.

To extract $\alpha_s$ and the heavy quark masses from $R_n$,
continuum extrapolation needs to be performed.
Since tree-level lattice artifacts cancel out in the reduced moments $R_n$, we expect
that discretization errors are proportional to $\alpha_s^n (a m_{h0})^j$. Therefore,
we fitted the lattice spacing dependence of $R_4$, $R_n/m_{h0}$, $n\ge 6$, and 
of the ratios $R_6/R_8$ and $R_8/R_{10}$ with the form
\begin{equation}
\sum_{n=1}^{N} \sum_{j=1}^{J} c_{nj} \alpha_s^{n} (a m_{h0})^{2j}.
\label{fit}
\end{equation}
Here for $\alpha_s$ we use the boosted lattice coupling defined as
\begin{equation}
\alpha_s^b(1/a) = \frac{1}{4\pi}\frac{g_0^2}{u_0^4} \, ,
\end{equation}
where $g_0^2=10/\beta$ is a bare lattice gauge coupling and $u_0$ is an averaged link value 
defined by the plaquette $u_0^4 = \langle {\rm Tr}U_\square\rangle/3$.
We use data corresponding
to $a m_{h0}<1.1$ to avoid uncontrolled cutoff effects. We note that the radius of convergence
of the Taylor series in  $a m_{h0}$ for the free theory is $\pi/2$ \cite{Bazavov:2017lyh}, and
thus, our upper limit on $a m_{h0}$ is well within the radius of convergence of the expansion.
The number of terms in Eq. (\ref{fit}) that have to be included when performing the continuum
extrapolations depends on the range of lattice spacings used in the fit. Restricting the fits
to small lattice spacings allows us to perform continuum extrapolations with fewer terms in
Eq. (\ref{fit}). Therefore, our general strategy for estimating continuum results was first to perform
the fits only using data at the smallest lattice spacings and few terms in Eq. (\ref{fit}), then perform
fits in an extended range of lattice spacings and more terms in Eq. (\ref{fit}), and finally compare
different fits to check for systematic effects. It also turned out that different quantities required
different numbers of terms in Eq. (\ref{fit}). The details of continuum extrapolations for different
quantities are given in Appendix B. Below we summarize the key features of the continuum extrapolations
of different moments and their ratios.

\begin{figure}[tb]
\begin{center}
\includegraphics[width=8.1cm]{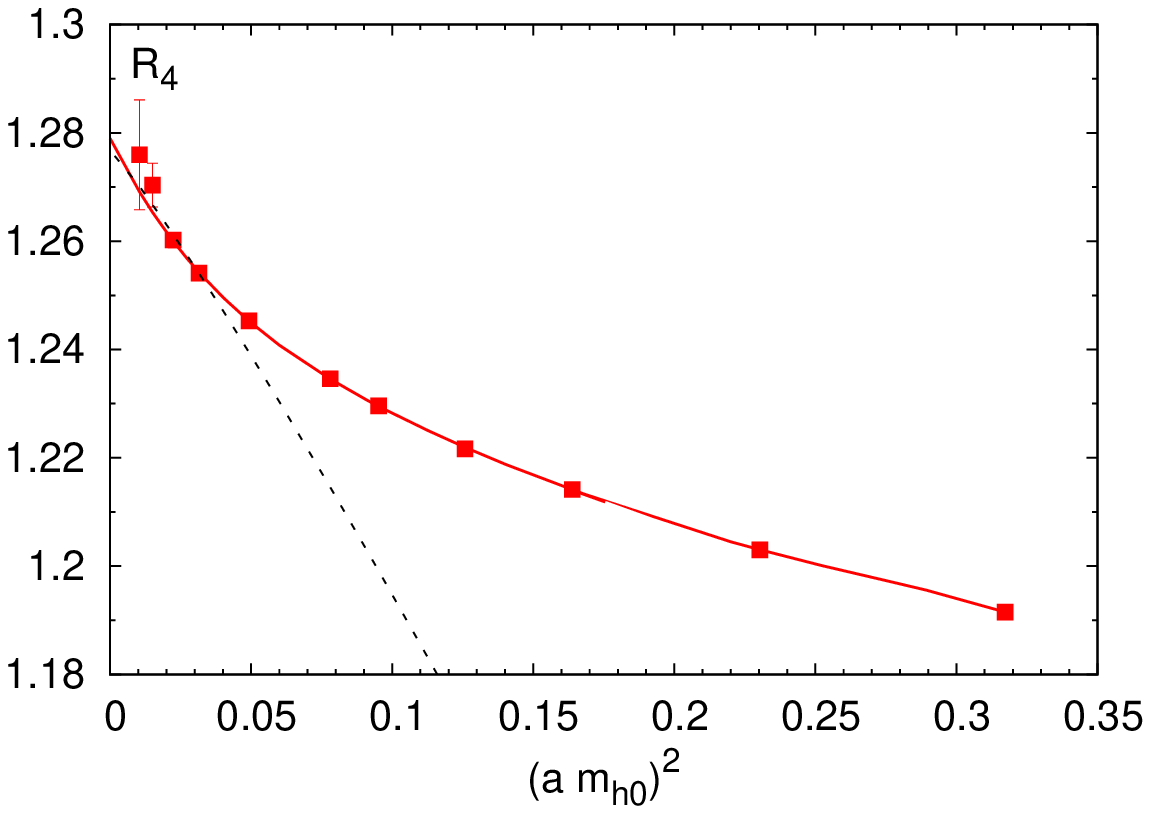} 
\includegraphics[width=8.1cm]{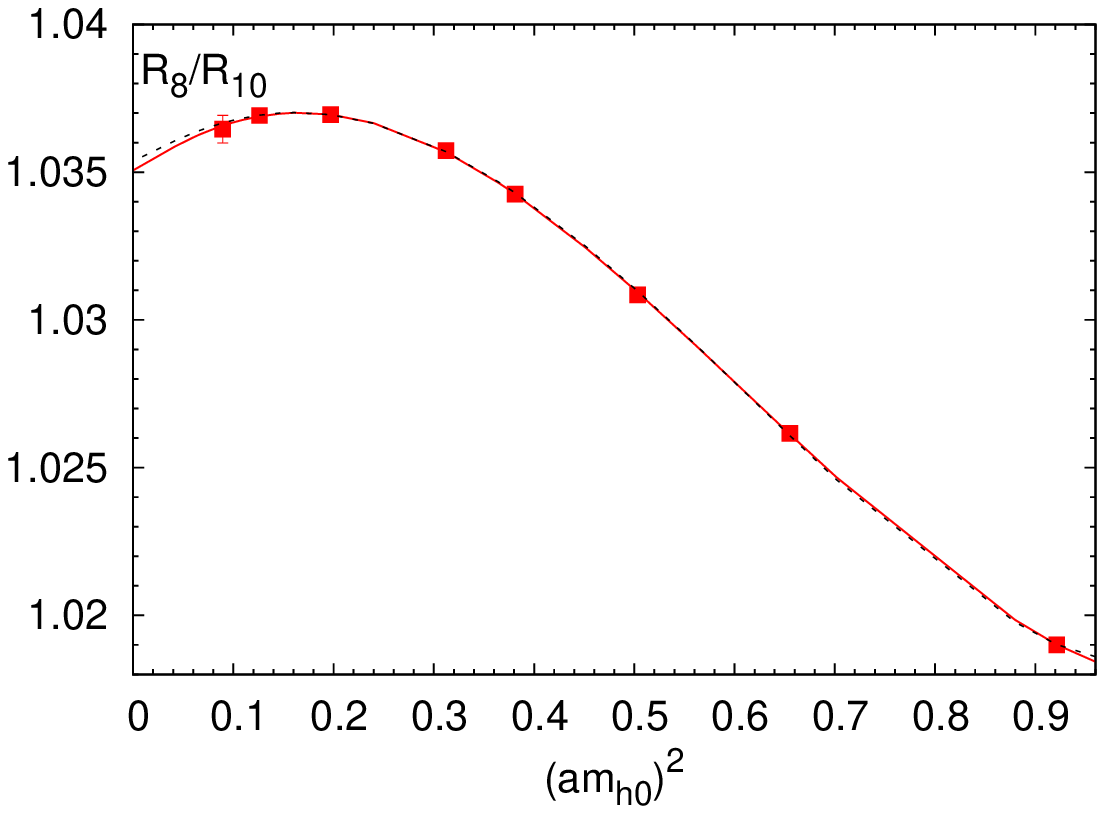}
\end{center}
\caption{Left: the lattice spacing dependence of $R_4$ for $m_h=m_c$ with the solid line showing
the fit to the 5th order polynomial in $(a m_h)^2$ with $N=2$, and the dashed line showing the $a^2$ fit. 
Right: the lattice spacing dependence of $R_8/R_{10}$ for $m_h=2 m_c$ together with different fits.
}
\label{fig:r4_rat8}
\end{figure}

\begin{figure}[tb]
\begin{center}
\includegraphics[width=8.1cm]{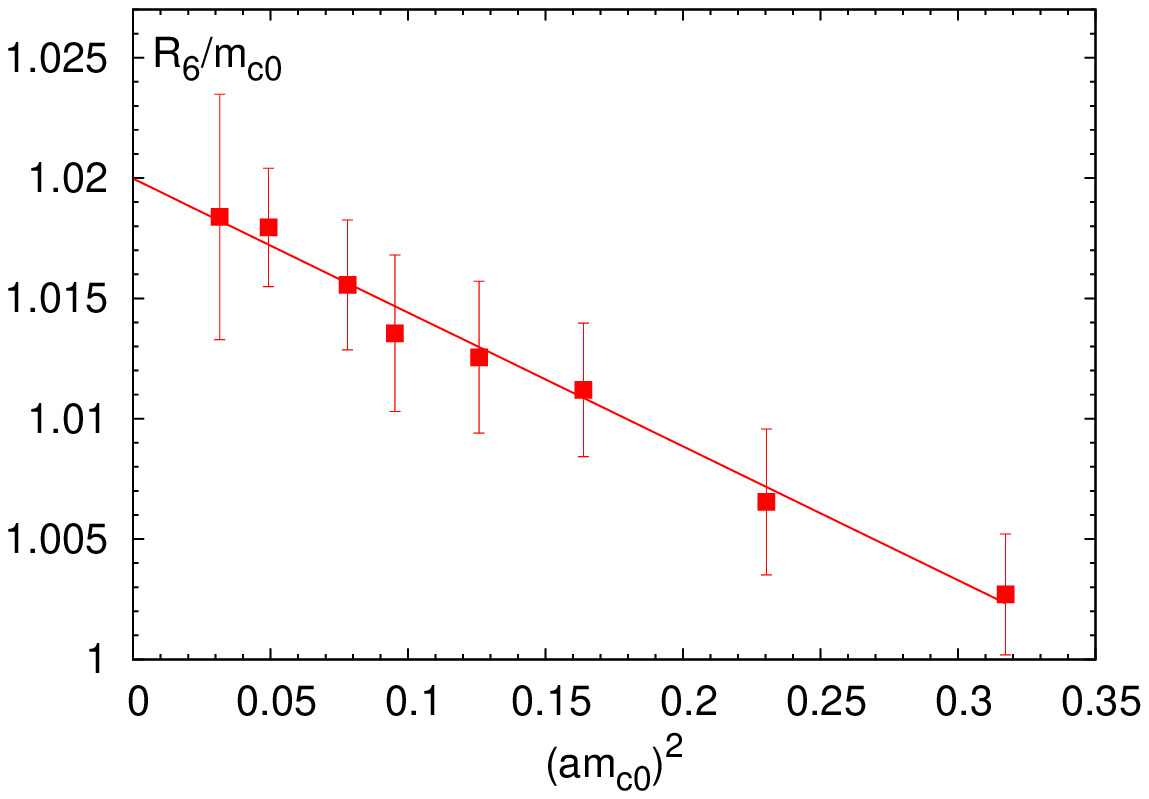}
\includegraphics[width=8.1cm]{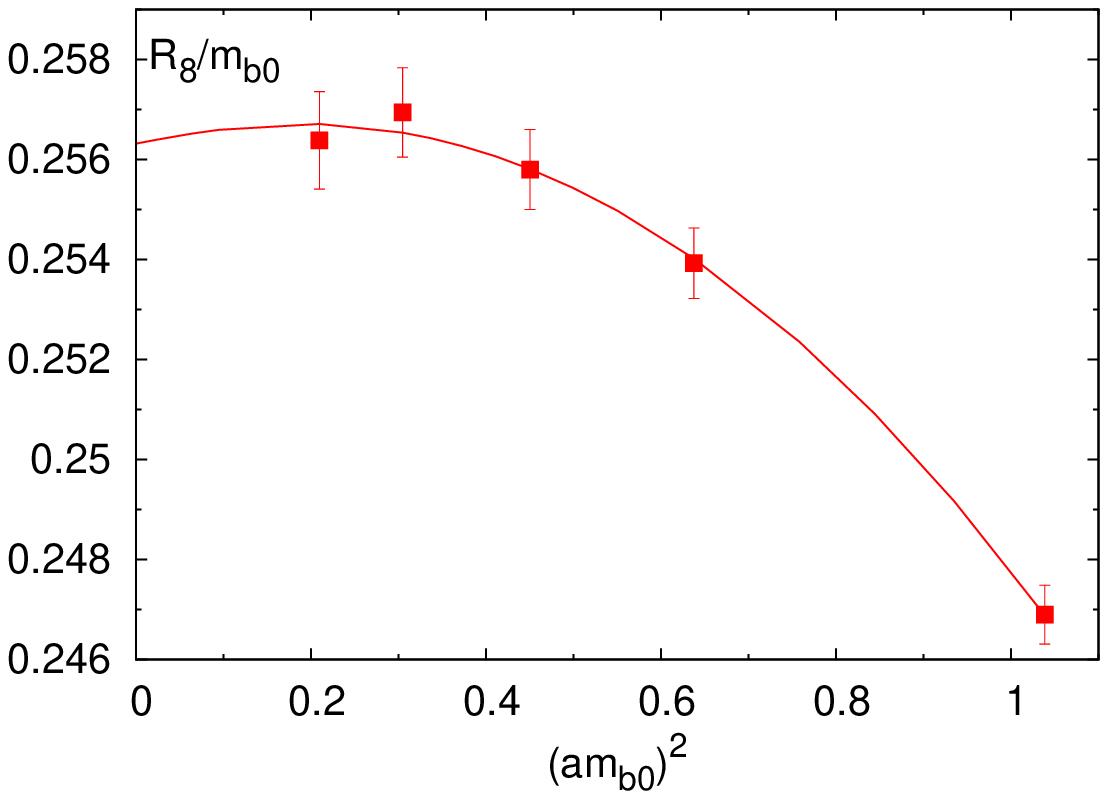}
\end{center}
\caption{Left: the lattice results for $R_6/m_{h0}$ for $m_h=m_c$ at
different lattice spacings. Also shown is the $a^2$
continuum extrapolation. Right: the lattice results for $R_8/m_{h0}$ for $m_h=m_b$
together with the $a^2+a^4$ fit.
}
\label{fig:r6_r8}
\end{figure}

The lattice spacing (cutoff) dependence
of $R_4$ turned out to be the most complicated. 
This is not completely surprising, as $R_4$ being the lowest moment, is most sensitive to short
distance physics.
Here we had to use up to fifth order polynomial in
$(a m_{h0})^2$ and at least two powers of $\alpha_s$ to describe the lattice data on $R_4$. Simpler fit forms only 
worked for the lowest mass and a very small value of the lattice spacings. For the 
ratios $R_6/R_8$ and $R_8/R_{10}$ we also had to use high order polynomials in $(a m_{h0})^2$,
though the leading order in $\alpha_s$ turned out to be sufficient. On the other hand, the lattice
spacing dependence of the ratios
$R_n/m_{h0}$ are described by the leading order ($\alpha_s a^2$) form or the leading order plus next-to-leading order 
($\alpha_s (a^2+a^4)$) form even for large values of $m_h$. 
To demonstrate these features we show
sample continuum extrapolations for the moments in Fig. \ref{fig:r4_rat8} and Fig. \ref{fig:r6_r8}.
In Fig. \ref{fig:r4_rat8} we show the results for $R_4$ at $m_h=m_c$
and $R_8/R_{10}$ for $m_h=2m_c$. As one can see from the left panel of the figure
the slope of the $a^2$ dependence of $R_4$ increases with decreasing
$a^2$. Therefore, if there are no data points at small $a$, the continuum limit may be underestimated.
The leading order fit only works for the four smallest lattice spacings but
agrees with the fit that uses a fifth order polynomial in $(a m_{h0})^2$ with $N=2$, 
and extends to the whole range of the lattice
spacings. Thus, the additional 
three lattice spacings included in this study are important
for cross-checking the validity of the continuum extrapolation, although
the correct continuum result for $R_4$ can be obtained without these additional data points.
This is important since finite volume effects are quite large for the three finest lattices.
For the $a$-dependence of the ratio $R_8/R_{10}$ we see the
opposite trend; the slope decreases at small $a$. 
Not having lattice results at small lattice spacing may lead to an overestimated
continuum result. 
In the right panel of Fig. \ref{fig:r4_rat8}, we show
the fits using fourth (solid line) and third order (dashed line) polynomials in $(a m_{h0})^2$
and leading order in $\alpha_s$. We see that the two fits give very similar results. Furthermore, we
find that higher order terms in $\alpha_s$ do no have a big impact here.
The $a$-dependence of $R_6/R_8$ was found to be similar.
The observed difference in the lattice spacing dependence of $R_4$ and the ratios $R_6/R_8$ and
$R_8/R_{10}$ as well as the difference in the systematic effects in the continuum extrapolations
will turn out to be important for cross-checking the consistency of the strong coupling constant
determination.
Because high order polynomials are needed for extrapolations when $a m_{h0}$ is large, continuum results for 
$R_4$, $R_6/R_8$ and $R_8/R_{10}$ could only be obtained for $m_h \le 3 m_c$.
For larger values of the quark masses we simply do not have enough data satisfying $a m_{h0}<1.1$ 
to perform the continuum extrapolations.

The lattice spacing dependence of $R_n/m_{h0},~n\ge 6$ turned out to be simpler and is well described by the next-to-leading order form for
all quark masses as can be seen for example in the right panel of Fig. \ref{fig:r6_r8}. For $R_6/m_{c0}$
even leading $a^2$ dependence is sufficient to describe the data, cf. the left panel of Fig. \ref{fig:r6_r8}.
Including higher order terms in $\alpha_s$ has no effect in this case.
\begin{table}
\begin{tabular}{|c|lll|}
\hline
$m_h$    &    $R_4$     &   $R_6/R_8$  &   $R_8/R_{10}$  \\
\hline
$1.0m_c$ &    1.279(4)  &   1.1092(6)  &     1.0485(8)   \\
$1.5m_c$ &    1.228(2)  &   1.0895(11) &     1.0403(10)  \\
$2.0m_c$ &    1.194(2)  &   1.0791(7)  &     1.0353(5)   \\
$3.0m_c$ &    1.158(6)  &   1.0693(10) &     1.0302(5)   \\
\hline
\end{tabular}
\caption{Continuum results for $R_4$, $R_6/R_8$, and $R_8/R_{10}$ at
different quark masses, $m_h$.}
\label{tab:contr4}
\end{table}

To obtain the continuum result 
for each quantity of interest, we performed many continuum extrapolations using different
ranges in the lattice spacing and different fit forms. We only consider fits that have 
$\chi^2/df$ of around one or smaller and take a weighted average of the corresponding
results to obtain the final continuum value. We use the scattering of different fits
around this averaged value to estimate the error of our continuum result. 
When the scattering in the central value of different fits around the average is considerably
smaller than the errors of the individual fits we take the typical errors of the fits
as our final error estimate.
In Appendix~\ref{app:B}
we give the details of this procedure. Our continuum results for $R_4$, $R_6/R_8$ and $R_8/R_{10}$
are shown in Table \ref{tab:contr4}. In Table \ref{tab:contr6} we give our continuum results
for $R_n/m_{h0}$ with $n\ge 6$. These two tables represent the main result of this study.

\begin{figure}
\includegraphics[width=5cm]{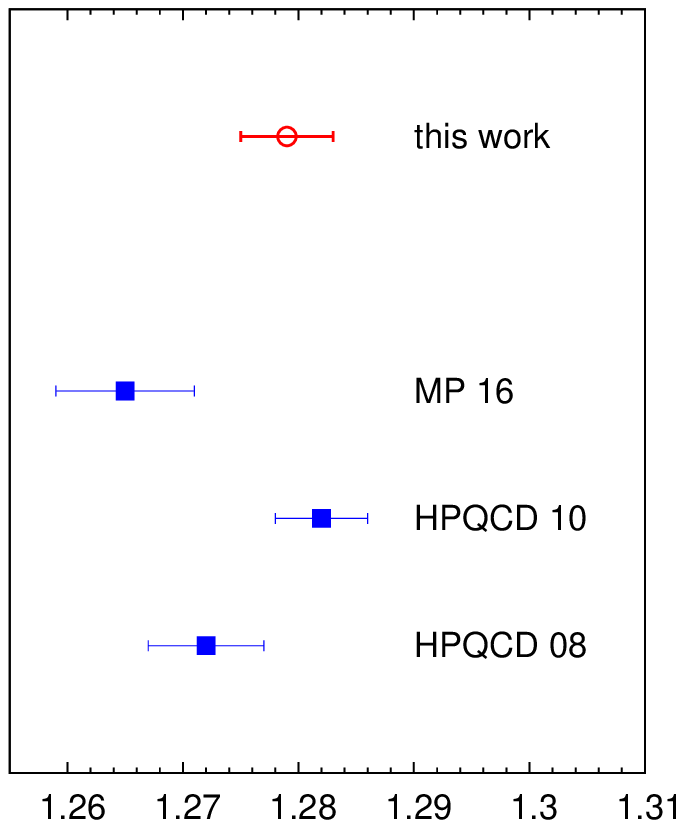}
\includegraphics[width=5cm]{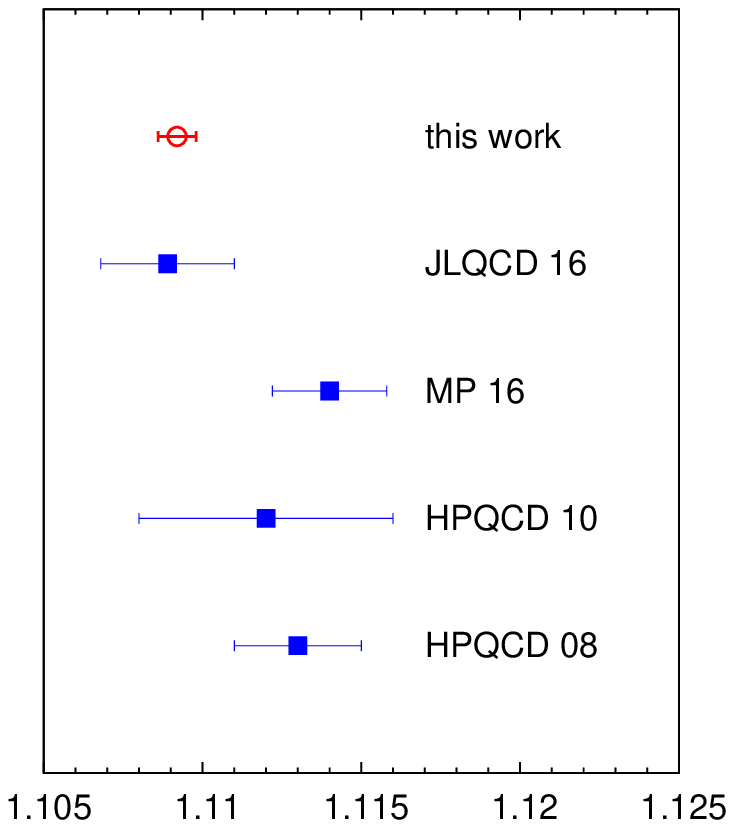}
\includegraphics[width=5cm]{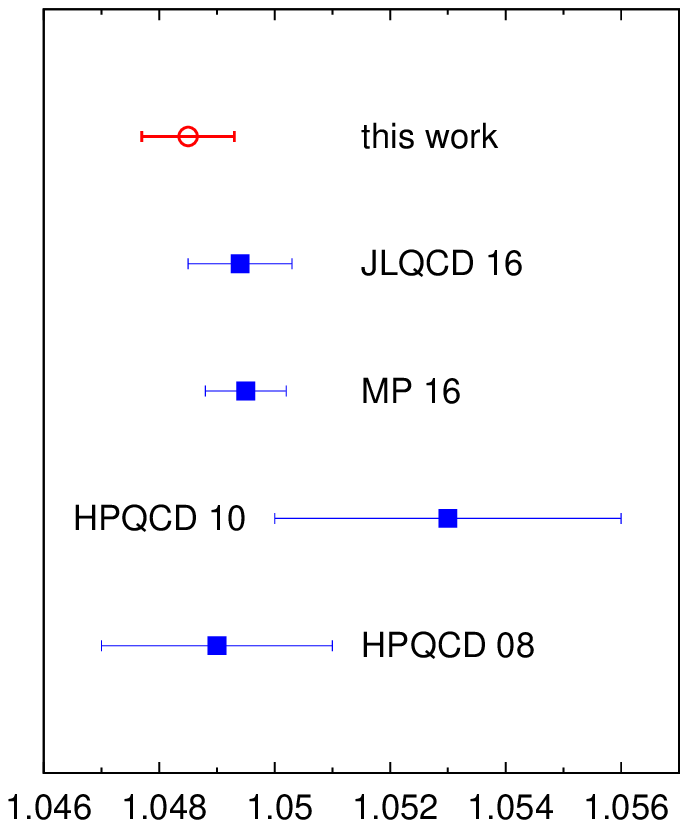}
\caption{Comparison of different lattice results for $R_4$ (left), $R_6/R_8$ (center) and $R_8/R_{10}$ (right); see the text for details.
The error on $R_6/R_8$ and $R_8/R_{10}$ for HPQCD 10 was obtained by propagating the errors on $R_6,~R_8$ and $R_{10}$ from Ref. \cite{McNeile:2010ji}.
}
\label{fig:compr4}
\end{figure}

As will be discussed in the following section using the continuum results on
the reduced moments $R_n$ and their ratios presented in Tables \ref{tab:contr4} and \ref{tab:contr6} 
one can obtain the strong
coupling constant as well as the values of the heavy quark masses, and may perform many important
cross-checks. However, before discussing the determination of $\alpha_s$ and the quark masses
let us compare our continuum results for the moments and their ratios with other lattice
determinations.
\begin{table}
\begin{tabular}{|c|lll|}
\hline
$m_h$    & $R_6/m_{h0}$ & $R_8/m_{h0}$ & $R_{10}/m_{h0}$ \\
\hline
$1.0m_c$ & 1.0195(20)   & 0.9174(20)   & 0.8787(50) \\
$1.5m_c$ & 0.7203(35)   & 0.6586(16)   & 0.6324(13) \\
$2.0m_c$ & 0.5584(35)   & 0.5156(17)   & 0.4972(17) \\
$3.0m_c$ & 0.3916(23)   & 0.3647(19)   & 0.3527(20) \\
$4.0m_c$ & 0.3055(23)   & 0.2859(12)   & 0.2771(23) \\
$m_b$    & 0.2733(17)   & 0.2567(17)   & 0.2499(16) \\
\hline
\end{tabular}
\caption{Continuum results for $R_n/m_{h0}$, $n \ge 6$ at
different quark masses, $m_h$.}
\label{tab:contr6}
\end{table}
In Fig. \ref{fig:compr4} we show the comparison of our continuum results on $R_4$, $R_6/R_8$, and $R_8/R_{10}$, which
can be used for $\alpha_s$ determination, 
with other lattice calculations for $m_h=m_c$. Our result on $R_4$ agrees with HPQCD results, published in 2008 \cite{Allison:2008xk}
and 2010 \cite{McNeile:2010ji} and labeled as HPQCD 08 and HPQCD 10, 
but is higher than the continuum
result from Ref. \cite{Maezawa:2016vgv}, denoted as MP 16. This is due to the fact that in Ref. \cite{Maezawa:2016vgv} simple
$a^2$ and $a^2+a^4$ continuum extrapolations, which cannot capture the correct dependence
on the lattice spacing as we now understand, have been used. The statistical errors on $R_4$ in those calculations were much
larger, and the inadequacy of simple $a^2$ and $a^2+a^4$ extrapolations was not apparent. 
Our result for
$R_6/R_8$ agrees with the JLQCD determination \cite{Nakayama:2016atf} (JLQCD 16) as well as with the HPQCD results 
published in 2008 and 2010 (labeled as HPQCD 08 and HPQCD 10).
However, the present continuum result for $R_6/R_8$ is smaller than the MP 16 result \cite{Maezawa:2016vgv}.
The reason for this is twofold. First, no lattice results for $a<0.04$ fm were available in Ref. \cite{Maezawa:2016vgv}.
As discussed above not having data for small enough $a$ may lead to an overestimated continuum limit for $R_6/R_8$. 
Second, in Ref. \cite{Maezawa:2016vgv} the continuum extrapolations
were performed using the simplest $a^2$ form with lattice results limited to $\beta<7.373$. Because of much larger statistical
errors, this fit was acceptable. However, with the new extended and more precise data a simple $a^2$ continuum extrapolation is no longer
appropriate, and the presence of the $a^4$ term leads to a smaller continuum result.
Finally for $R_8/R_{10}$ all lattice results agree within errors.
\begin{figure}
\includegraphics[width=5cm]{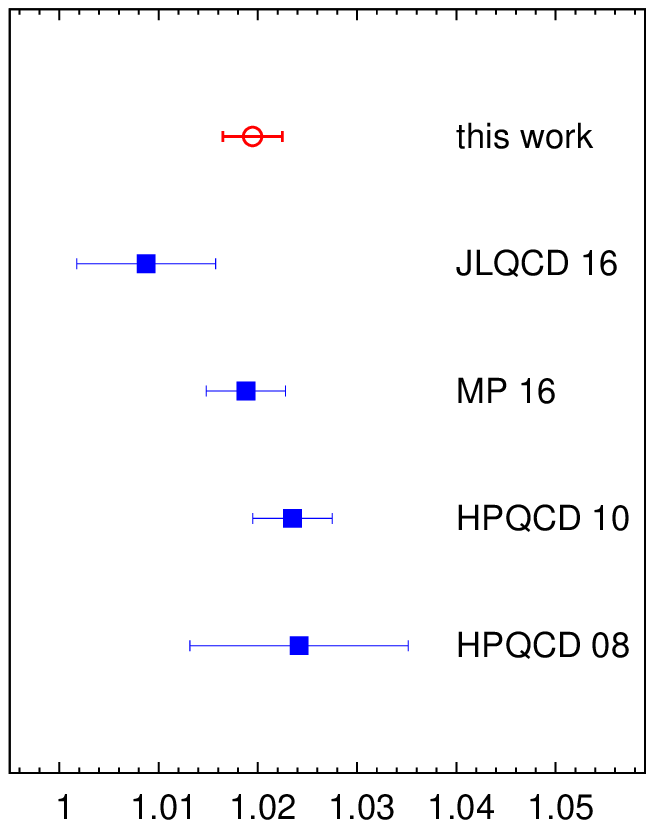}
\includegraphics[width=5cm]{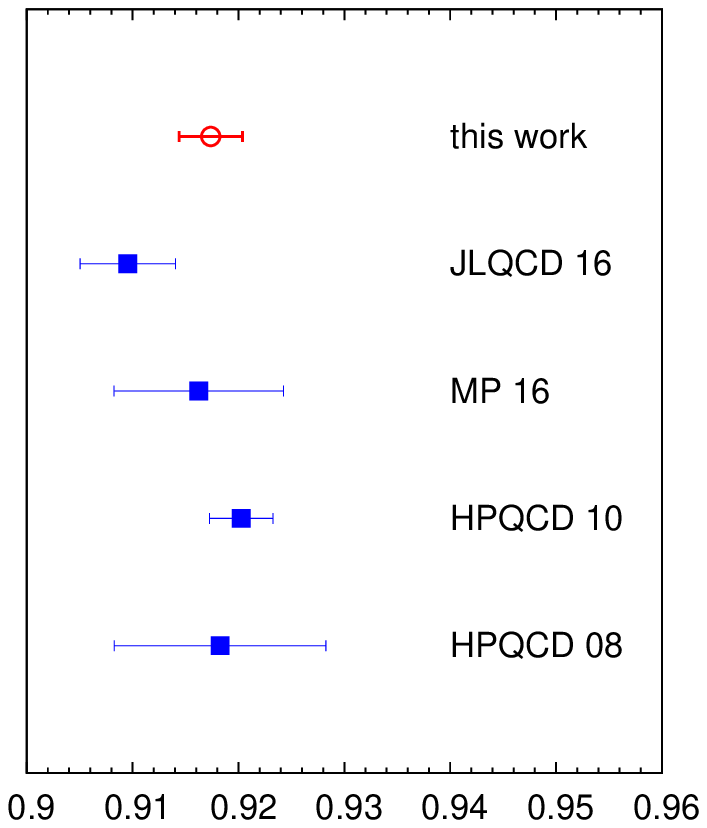}
\includegraphics[width=5cm]{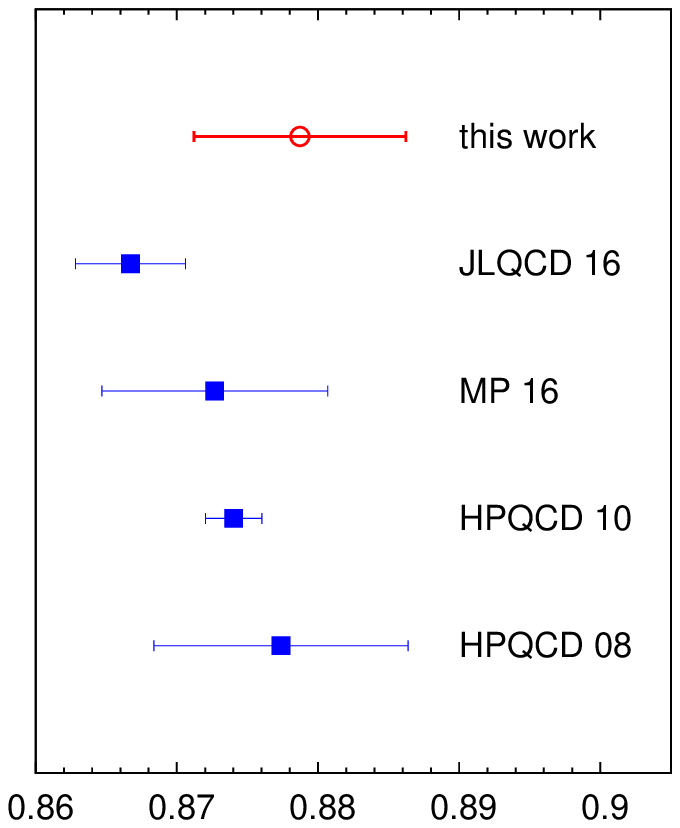}
\caption{Comparison of different lattice results for $R_6$ (left), $R_8$ (center), and $R_{10}$ (right); see the text for details.}
\label{fig:compr6}
\end{figure}
In Figure  \ref{fig:compr6} we compare our continuum results on $R_6$, $R_8$, and $R_{10}$ for $m_h=m_c$ with other lattice studies,
including the work by JLQCD \cite{Nakayama:2016atf}, Maezawa and Petreczky \cite{Maezawa:2016vgv}, and HPQCD \cite{Allison:2008xk,McNeile:2010ji}.
As one can see from the figure our results agree with other lattice works within errors. Perhaps, this is not too surprising
as the $a$-dependence of these reduced moments is well described by a simple $\alpha_s (a^2+a^4)$ form.

\section{Strong coupling constant and heavy quark masses}
\label{sec:asmh}

From the continuum results on $R_4$ and the ratios of the reduced moments, $R_6/R_8$, and $R_8/R_{10}$, we can extract 
the value of the strong coupling constant. As discussed in the previous section we choose the renormalization
scale to be $\mu=m_h(\mu_m=m_h)$ and solve the nonlinear equations to obtain the value of $\alpha_s(\mu=m_h)$.
To estimate the error due to the truncation of the perturbative series in $R_n$ we assume that the coefficient
of the unknown $\alpha_s^4$ term varies between $-5r_{n3}$ and $+5r_{n3}$, i.e. $r_{n4}=\pm 5\times r_{n3}$. 
This error estimate should be sufficiently conservative.
We also take into account the non-perturbative contribution
to the reduced moments at tree level according to Ref. \cite{Broadhurst:1994qj} and use the value of the gluon condensate
given in Eq. (\ref{cond}). In Table \ref{tab:as}, we give our results for $\alpha_s(\mu=m_h)$ for different quark masses.
We see that both the perturbative uncertainty as well as the uncertainty due to the gluon condensate drastically
decrease with increasing $m_h$. The $\alpha_s$ values determined from $R_6/R_8$ and $R_8/R_{10}$ have much
larger perturbative uncertainties than the ones from $R_4$.
There is a slight tension between the values of $\alpha_s$ determined from $R_4$ and $R_6/R_8$, and
the values obtained from $R_8/R_{10}$ for $m_h=m_c,~1.5m_c$. 
The strong coupling constant determined from $R_8/R_{10}$ is lower for these $m_h$ values.
A similar trend was observed in the 2008 HPQCD analysis \cite{Allison:2008xk}. 
For $m_h=2m_c$, we find that all three determinations of $\alpha_s$ from $R_4$, $R_6/R_8$
and $R_8/R_{10}$ agree within errors.
To obtain our final estimate
of $\alpha_s(\mu)$ for $\mu=m_c,~1.5m_c,~2m_c$ and $3m_c$ we performed a weighted average of the results
obtained from $R_4$, $R_6/R_8$, and $R_8/R_{10}$, which is justified since the systematic errors
on $\alpha_s$ obtained from these quantities are largely uncorrelated, while statistical
errors are negligible. There is some correlation in the error due to the gluon condensate since
the error of the gluon condensate enters in all of the quantities. However, performing the weighted
average without the error due to the gluon condensate only leads to very small changes in $\alpha_s(m_h)$, if any.
The average values of $\alpha_s(m_h)$ are given in the fifth column of Table \ref{tab:as}.
The uncertainty of the averaged $\alpha_s$ values was determined
such that it agrees with all individual $\alpha_s$ extraction within the estimated errors.
\begin{table}
\begin{tabular}{|c|lllll|}
\hline
$m_h$    & $R_4$              & $R_6/R_8$           &  $R_8/R_{10}$        & av.         & $\Lambda_{\overline{MS}}^{n_f=3}$ MeV \\
\hline
$1.0m_c$ & 0.3815(55)(30)(22) & 0.3837(25)(180)(40) &  0.3550(63)(140)(88) & 0.3782(65)  & 314(10)  \\
$1.5m_c$ & 0.3119(28)(4)(4)   & 0.3073(42)(63)(7)   &  0.2954(75)(60)(17)  & 0.3099(48)  & 310(10) \\
$2.0m_c$ & 0.2651(28)(7)(1)   & 0.2689(26)(35)(2)   &  0.2587(37)(34)(6)   & 0.2648(29)  & 284(8)  \\
$3.0m_c$ & 0.2155(83)(3)(1)   & 0.2338(35)(19)(1)   &  0.2215(367)(17)(1)  & 0.2303(150) & 284(48) \\
\hline
\end{tabular}
\caption{The values of $\alpha_s(\mu=m_h)$ for different heavy quark masses, $m_h$, extracted 
from $R_4$, $R_6/R_8$, and $R_8/R_{10}$. 
The first, second, and third errors correspond to the lattice error, the perturbative truncation error, and the error
due to the gluon condensate.
In the fifth column, the averaged value of $\alpha_s$ is shown (see the text).
The last column gives the value of $\Lambda_{\overline{MS}}^{n_f=3}$ in MeV.}
\label{tab:as}
\end{table}

Using the continuum results for $R_n/m_{h0},~n\ge 6$,  given in Table \ref{tab:contr6}, together with the corresponding perturbative
expression for $R_n,~n\ge 6$, and the averaged value of $\alpha_s(m_h)$ given in the fifth column of Table \ref{tab:as}, we obtain
the values of $m_h$ in the $\overline{MS}$ scheme at $\mu=m_h$. These are presented in Table \ref{tab:mh}.
The differences in the central values of the heavy quark masses obtained from $R_6$, $R_8$, and $R_{10}$ are much smaller
than the estimated errors.
Therefore, we calculated the corresponding average to obtain our final estimates
and for the heavy quark masses. Similarly, the error estimates were obtained as averages
over the error estimates obtained from $R_6$, $R_8$, and $R_{10}$.
The results are given in the last column of Table \ref{tab:mh}.
The errors on the heavy quark masses in the table do not contain the overall scale error yet.
\begin{table}
\begin{tabular}{|l|cccc|}
\hline
$m_h$     &   $R_6$                   &       $R_8$               &         $R_{10}$          &   av.       \\ 
\hline
$1.0m_c$  &  1.2740(25)(17)(11)(61)   &  1.2783(28)(23)(00)(43)   &  1.2700(72)(46)(13)(33)   & 1.2741(42)(29)(8)(46)  \\
$1.5m_c$  &  1.7147(83)(11)(03)(60)   &  1.7204(42)(14)(00)(40)   &  1.7192(35)(29)(04)(30)   & 1.7181(53)(18)(2)(43)  \\
$2.0m_c$  &  2.1412(134)(07)(01)(44)  &  2.1512(71)(10)(00)(29)   &  2.1531(74)(19)(02)(21)   & 2.1481(93)(12)(1)(31)  \\
$3.0m_c$  &  2.9788(175)(06)(00)(319) &  2.9940(156)(08)(00)(201) &  3.0016(170)(16)(00)(143) & 2.9915(167)(10)(0)(220) \\
$4.0m_c$  &  3.7770(284)(06)(00)(109) &  3.7934(159)(08)(00)(68)  &  3.8025(152)(15)(00)(47)  & 3.7910(198)(10)(0)(75)  \\
$m_b$     &  4.1888(260)(05)(00)(111) &  4.2045(280)(07)(00)(69)  &  4.2023(270)(14)(00)(47)  & 4.1985(270)(9)(0)(76) \\
\hline
\end{tabular}
\caption{The heavy quark masses in the $\overline{MS}$ scheme at $\mu=m_h$ in GeV for different values of $m_h$.
The first, second, third, and fourth errors correspond 
to the error of the lattice result, the perturbative truncation error, the error due
to the gluon condensate, and the error from $\alpha_s$, respectively.
The last column shows the average of the masses determined from $R_6$, $R_8$, and $R_{10}$.}
\label{tab:mh}
\end{table}

Combining the information from the above table with the value of $\alpha_s(m_h)$ in the fifth column
of Table \ref{tab:as}, we can obtain the values of $\Lambda_{\overline{MS}}^{n_f=3}$, which are
given in the last column of Table \ref{tab:as}. 
To obtain the $\Lambda_{\overline{MS}}^{n_f=3}$ from the value of the coupling
at $\mu=m_h$, we use the implicit scheme given by Eq. (5) of Ref. \cite{Chetyrkin:2000yt}. 
We also calculated the $\Lambda_{\overline{MS}}^{n_f=3}$ in the explicit
scheme given by Eq. (4) of Ref. \cite{Chetyrkin:2000yt}, and the small differences between the two schemes have been treated as systematic errors. Finally, we included
the error in the scale determination in the values of $m_h$ and the error in $\alpha_s(m_h)$. All these errors have
been added in quadrature. We see that the value of $\Lambda_{\overline{MS}}^{n_f=3}$ determined
from the $m_h=2 m_c$ data is $2.5 \sigma $ lower that the ones obtained from the $m_h=m_c$ and $m_h=1.5 m_c$ data.
To obtain our final estimate for $\Lambda_{\overline{MS}}^{n_f=3}$, we take
an (unweighted) average of the data in the last column of Table \ref{tab:as} and use the spread around this central value 
as our (systematic) error:
\begin{equation}
\Lambda_{\overline{MS}}^{n_f=3}=298 \pm 16~{\rm MeV}.
\label{Lam}
\end{equation}
The too low value of $\Lambda_{\overline{MS}}^{n_f=3}$ obtained for $m_h=2 m_c$ is of some concern. One could imagine that the continuum
extrapolation of $R_4$ at this quark mass is not reliable, and the corresponding $\alpha_s$ should not be considered
in the analysis. 
On the other hand, the continuum extrapolation of $R_6/R_8$ and $R_8/R_{10}$ is more robust, and 
any systematic effect due to coarse lattices will overestimate the continuum
limit and  make $\alpha_s$ larger.
If we determine $\alpha_s(2m_c)$ using only the results for $R_6/R_8$ and $R_8/R_{10}$, we obtain
a value for $\alpha_s(2m_c)$, which is only one tenth sigma different from the value in Table \ref{tab:as}.
Finally, if we take $\alpha_s(2m_c)$ and $\alpha_s(3m_c)$ only from $R_6/R_8$ we obtain $\Lambda_{\overline{MS}}^{n_f=3}=293(10)$ MeV
and $\Lambda_{\overline{MS}}^{n_f=3}=293(12)$ MeV, respectively, 
resulting in an  average of $\Lambda_{\overline{MS}}^{n_f=3}=302(11)$ MeV, which lies well within the uncertainty of
the above result.

Using Eq. (\ref{Lam}) for $\Lambda_{\overline{MS}}^{n_f=3}$, we can calculate $\alpha_s(\mu=4 m_c)$, and $\alpha_s(\mu=m_b)$
and also determine the corresponding quark masses,
$m_h=4m_c$ and $m_h=m_b$. These are presented in the last two rows of Table \ref{tab:mh}. Again,
we see that the differences in the heavy quark masses obtained from $R_6$, $R_8$, and $R_{10}$ are smaller than the estimated
errors, suggesting
that the quark mass determination from the reduced moments is under control even for the largest values of the heavy  quark masses.
To obtain the final value of the quark masses, we use the same procedure as before.

\begin{figure}
\includegraphics[width=8cm]{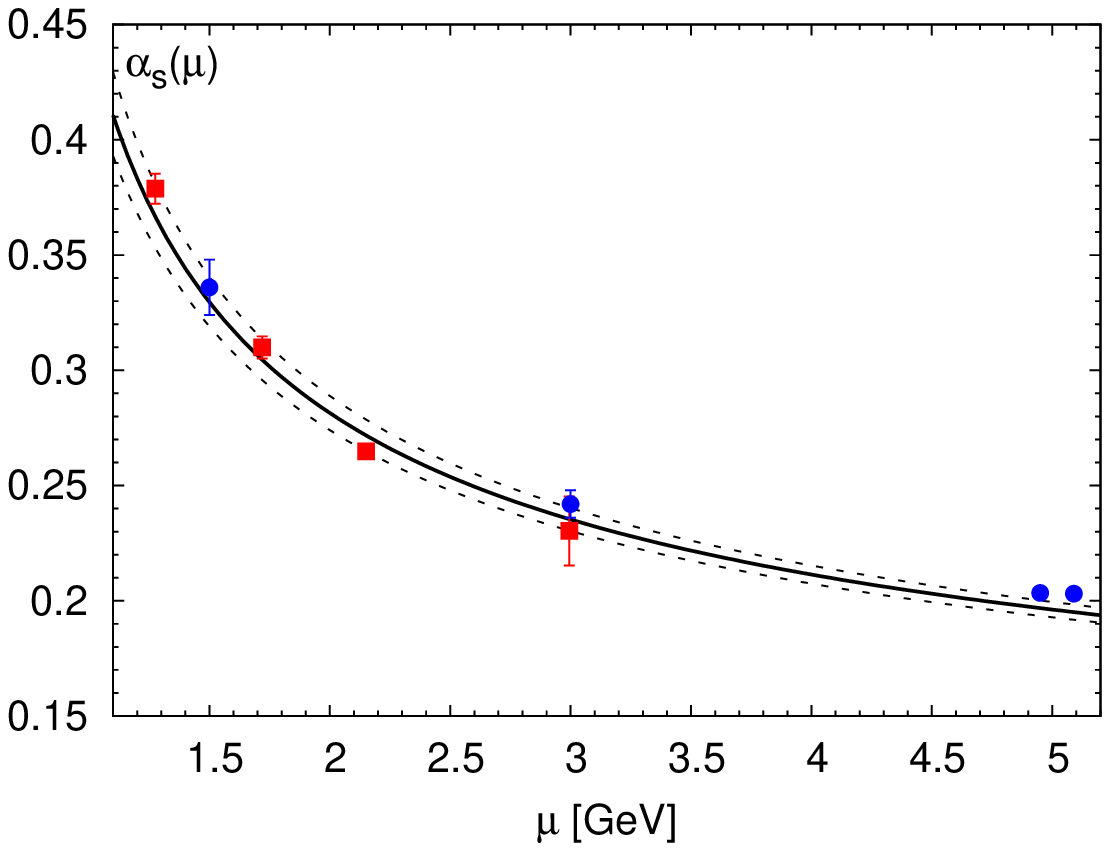}
\includegraphics[width=8cm]{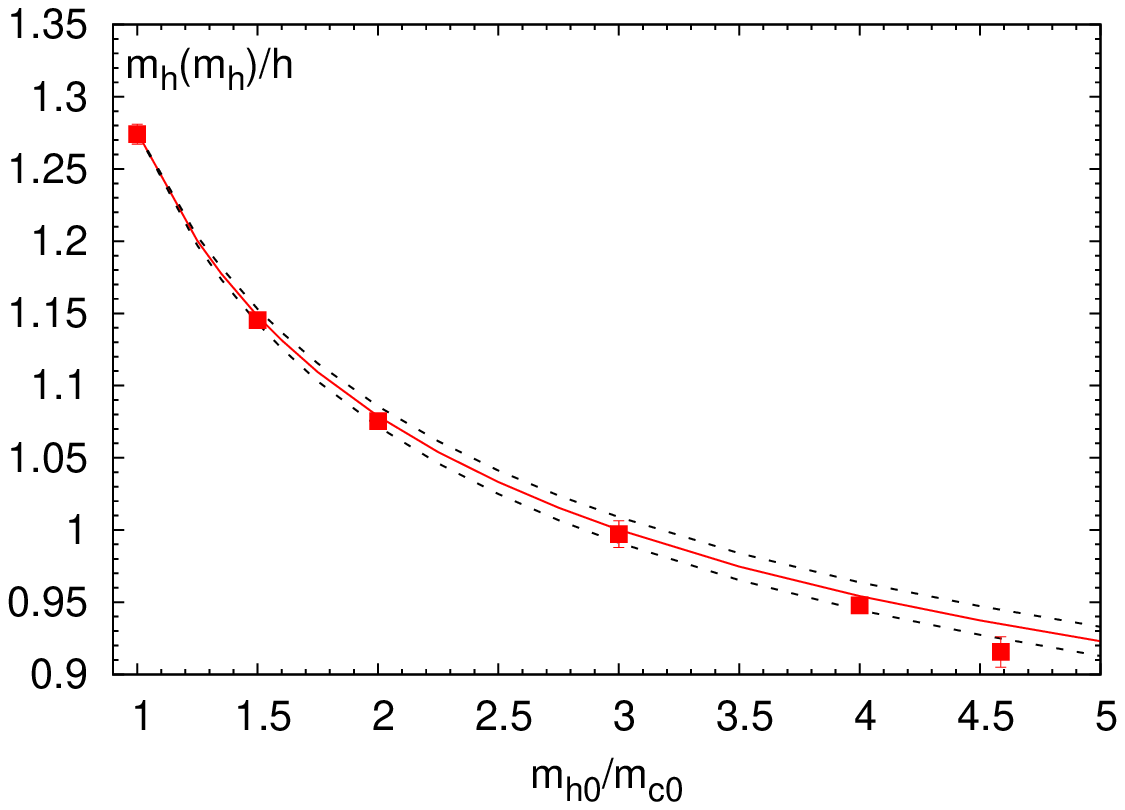}
\caption{Left: The running coupling constant in three-flavor QCD 
corresponding to $\Lambda_{\overline{MS}}^{n_f=3}=298(16)$ MeV.
The solid line corresponds to the central value, while the dashed lines correspond to the error band. The red
squares show the lattice results of this work (fifth column of Table \ref{tab:contr4}). 
The blue circles from left to right correspond to the determination of $\alpha_s$
for the static quark anti-quark potential \cite{Bazavov:2014soa} and from the moments of quarkonium correlators 
\cite{Allison:2008xk,McNeile:2010ji,Chakraborty:2014aca}. 
The result of Ref. \cite{McNeile:2010ji} has been shifted horizontally for better visibility.
Right: The running of the quark mass $m_h(m_h)/h,~h=m_h/m_c$. The line corresponds to four-loop running, while
the dashed lines correspond to the uncertainty of $\Lambda_{\overline{MS}}^{n_f=3}$. The data points are taken
from the fifth column of Table \ref{tab:mh} and in the rightmost data point, we also included the error on the ratio
$m_b/m_c$ from Eq. (\ref{mb_by_mc}).
}
\label{fig:running}
\end{figure}

With all the above information, we can now test the running of the strong coupling constant and the heavy quark masses.
To study the running of the heavy quark mass, we consider the ratio $m_h(m_h)/h$, where $h=m_h/m_c$. We can
think about this quantity as the charm quark mass at different scales.
The running of $\alpha_s$ and the running of the heavy quark mass are shown in Fig. \ref{fig:running}. In the figure we show the coupling constant determined in other
lattice studies, including the determination from the static quark anti-quark energy \cite{Bazavov:2014soa}
and moments of quarkonium correlators \cite{Allison:2008xk,McNeile:2010ji,Chakraborty:2014aca}. The results of Refs. \cite{Allison:2008xk,Chakraborty:2014aca}
correspond to the four-flavor theory. We converted the corresponding values of $\alpha_s$ to the three-flavor scheme using perturbative decoupling
as implemented in the RunDeC package \cite{Chetyrkin:2000yt} and assuming a charm threshold of $1.5$ GeV. There is fairly good agreement between the running coupling constant
in this study and other lattice determinations.
We can also see from Fig. \ref{fig:running} that the running of the heavy quark mass follows the expectation
very well.

We can convert our result on $\Lambda_{\overline{MS}}^{n_f=3}$ to $\alpha_s$ for $n_f=5$ at scale $\mu=M_Z$ 
by including the contribution of the charm and bottom quarks to the running of the coupling constant.
We do this by using the RunDeC package \cite{Chetyrkin:2000yt} and first match 
$\alpha_s$ at the charm threshold, which we choose to be $1.5$ GeV, and then 
at the bottom quark threshold, which we choose to be $4.7$ GeV, and finally evolve $\alpha_s$ 
to $\mu=M_Z$. We get 
\begin{equation}
\alpha_s (M_Z , n_f=5) = 0.1159(12) \, .
\label{asmz}
\end{equation}
This result agrees with $\alpha_s (M_Z , n_f=5) = 0.11622(83)$
obtained in Ref. \cite{Maezawa:2016vgv} that used the same lattice setup
within the error, but the error increased despite much smaller statistical errors and more lattice data
points. The use of many fit forms, several heavy quark masses and of the ratios of reduced moments resulted in a more conservative
error estimate. Let us now compare our results for $\alpha_s$ given in Eq. (\ref{asmz}) with other
determinations. 
Our result is smaller than the PDG average $\alpha_s(M_Z,n_f=5)=0.1181(11)$ \cite{PDG18} by $1.4 \sigma$, and it is
smaller than 
the FLAG average $\alpha_s(M_Z,n_f=5)=0.11823(81)$ \cite{FLAG19} by $1.6\sigma$.
It is also smaller that the determination of $\alpha_s(M_Z,n_f=5)=0.11852(84)$ by the ALPHA collaboration using the
Schr\"odinger functional method \cite{Bruno:2017gxd} by $1.8\sigma$.
Two recent lattice determinations, one from the combined analysis of $R_6$, $R_8$, and $R_{10}$ 
\cite{Nakayama:2016atf}, and another one from hadronic vacuum polarization \cite{Hudspith:2018bpz} give
values $\alpha_s(M_Z,n_f=5)=0.1176(26)$ and $\alpha_s(M_Z,n_f=5)=0.1181(27)^{+8}_{-22}$, respectively.
These agree with our results though the central values are higher. 
The very recent analysis of the static quark anti-quark energy resulted in 
$\alpha_s(M_Z,n_f=5)=0.1166^{+0.0010}_{-0.0011}(stat)^{+0.0018}_{-0.0017}(sys)$ and
$\alpha_s(M_Z,n_f=5)=0.1179\pm 0.0007(stat)^{+0.0013}_{-0.0012}(sys)$,
depending on the analysis strategy \cite{Takaura:2018lpw,Takaura:2018vcy}. 
These again agree with our result within errors.
Finally, a recent phenomenological 
estimate based on the bottomonium spectrum gave $\alpha_s(M_Z,n_f=5)=0.1178(51)$ \cite{Mateu:2017hlz}, 
which is again compatible
with our result. 

Now, let us compare the determination of the charm quark mass, $m_c$. Using the result
from Table \ref{tab:mh} and adding the scale uncertainty from $r_1$, we have 
\begin{equation}
m_c(\mu=m_c,n_f=3)=1.2741(101)  {\rm GeV}. 
\label{mc3f}
\end{equation}
This result agrees with the previous (2+1)-flavor determination of the charm quark mass using the
HISQ action, $m_c(\mu=m_c,n_f=3)=1.267(12)$ GeV \cite{Maezawa:2016vgv}. The charm quark mass
determination in Ref. \cite{Maezawa:2016vgv} was criticized in the FLAG review because of the small volumes and slightly larger
than physical light sea quark masses ($m_s/20$ instead of $m_s/27$), and as the result, the corresponding determination did
not enter the FLAG average \cite{FLAG19}. We would like to point out that this criticism was not fully justified.
As shown in the present analysis, the effect of the light sea quark mass in the range from
$m_s/20$ to $m_s/5$ is negligible, and thus the use of $m_l=m_s/20$ for the light quark masses
can hardly affect the charm quark mass determination. The lattice spacing dependence of $R_6/m_{c0}$ is
well described by the $a^2$ form in the entire range range of lattice spacings available. Therefore, the 
continuum extrapolation is not significantly affected by the data at small lattice spacings, where
the physical volume is small according to the FLAG criteria; see the discussion in Appendix B.

Using Eq. (\ref{mc3f}) and performing the matching to four flavor theory with RunDeC we
obtain
\begin{equation}
m_c(\mu=m_c,n_f=4)=1.265(10)~ {\rm GeV}.
\label{mc}
\end{equation}
This result agrees with the (2+1+1)-flavor determination by the
HPQCD Collaboration $m_c(\mu=m_c,n_f=4)=1.2715(95)$ \cite{Chakraborty:2014aca}, confirming the observation
that perturbative decoupling of the charm quark is justified \cite{Athenodorou:2018wpk}.
It is customary to quote the result for the charm quark mass at scale $\mu=3$ GeV.
Evolving our result for $m_c(m_c)$
to $\mu=3$ GeV and  with the RunDeC package we obtain
\begin{equation}
m_c(\mu=3~{\rm GeV},n_f=4)=1.001(16)~{\rm GeV}.
\label{mc3}
\end{equation}
As before, the matching to the four flavor theory was carried out at $1.5$ GeV. 
The uncertainty in $\Lambda_{\overline{MS}}^{n_f=3}$ given Eq. (\ref{Lam}) has a significant effect
on the evolution and thus leads to a larger error on $m_c$ at this scale.
Our result agrees well with the
HPQCD determinations that rely on the moments 
of quarkonium correlators, $m_c(\mu=3~{\rm GeV},n_f=4)=0.9851(63)~{\rm GeV}$ \cite{Chakraborty:2014aca}, 
and the HQPCD result based on the RI-SMOM scheme, 
$m_c(\mu=3~{\rm GeV},n_f=4)=0.9896(61)~{\rm GeV}$  \cite{Lytle:2018evc} as well
as with the Fermilab-MILC-TUMQCD result based on the MRS scheme, 
$m_c(\mu=3~{\rm GeV},n_f=4)=0.9843(56)~{\rm GeV}$ \cite{Bazavov:2018omf}.
Furthermore, we also agree with the value reported by the JLQCD collaboration, $m_c(\mu=3~{\rm GeV},n_f=4)=1.003(10)$ GeV.

Finally, we discuss the determination of the bottom quark mass, $m_b$.
Using the result from Table \ref{tab:mh} and taking into account the scale error we get
$m_b(\mu=m_b,n_f=3)=4.1985(371)$ GeV. We
estimate the bottom quark mass for five flavors by evolving the result with the RunDec package as before, which results in
\begin{equation}
m_b(\mu=m_b,n_f=5)=4.188(37)~{\rm GeV}.
\label{mb}
\end{equation}
The above error also includes the uncertainty in the value of $\Lambda_{\overline{MS}}^{n_f=3}$.
This value for $m_b$ is in good agreement with other lattice determinations by the HQPCD collaboration, 
$m_b(\mu=m_b,n_f=5)=4.162(48)~{\rm GeV}$ \cite{Chakraborty:2014aca},
by the ETMC collaboration, $m_b(\mu=m_b,n_f=5)=4.26(10)~{\rm GeV}$ \cite{Bussone:2016iua}, 
by the Fermilab-MILC-TUMQCD collaboration,
$m_b(\mu=m_b,n_f=5)=4.197(14)~{\rm GeV}$ \cite{Bazavov:2018omf}, 
as well as with the previous (2+1)-flavor determination using the HISQ action,
$m_b(\mu=m_b,n_f=5)=4.184(89)~{\rm GeV}$ \cite{Maezawa:2016vgv}.
Moreover, our result also agrees with the value $m_b(\mu=m_b,n_f=5)=4.216(39)$ GeV obtained from bottomonium
phenomenology \cite{Mateu:2017hlz}. We could have also determined 
the bottom quark mass from the value of $m_c(\mu=m_c)$ and the ratio
$m_b/m_c$ obtained in section \ref{sec:lat_setup}. 
As one can see from Fig. \ref{fig:running}, this would have resulted in a value
which is compatible with the above result but has a significantly larger error at scale $\mu=m_b$.

\section{Conclusion}
\label{sec:conclusion}

In this paper, we calculated the moments of quarkonium correlators for several heavy quark
masses in (2+1)-flavor QCD using the HISQ action. 
From the moments of quarkonium correlators we extracted the strong coupling constant and the 
heavy quark masses. 
Our main results are given in Tables \ref{tab:contr4} and \ref{tab:contr6} 
and by Eqs. (\ref{Lam}), (\ref{asmz}),
(\ref{mc}), and (\ref{mb}).
We improved and extended the previous (2+1)-flavor HISQ analysis published
in Ref. \cite{Maezawa:2016vgv}. We drastically reduced the statistical errors on the moments
by using random color
wall sources, extended the calculations to smaller lattice spacings and considered
several values of the heavy quark masses in the region between the charm and bottom quark mass.
The novel feature of our analysis is the use of very small lattice spacings, which enables
reliable continuum extrapolations. The use of the very small lattice spacings, however, comes
with small physical volumes. We showed that the use of small physical volumes is not a major
problem for the analysis of the moments of quarkonium correlators.

The calculations of the reduced moments at several heavy quark masses enabled us to map
out the running of the coupling constant at low energy scales.
It also allowed for an additional control of the systematic errors
due to the truncation of the perturbative series as the perturbative errors go down with
an increasing heavy quark mass. Comparison of $\alpha_s(m_h)$ determined
for different heavy quark masses, $m_h$, led to a more conservative error
estimate for the $\Lambda$-parameter compared to the estimate one would get
just using the results for $m_h=m_c$. It is clear that 
extending the perturbative calculations of the 
moments of quarkonium correlators to higher order will be very useful and is likely
to lead to a more precise determination of $\alpha_s$.

Evolving the low energy determination of $\alpha_s$
to $\mu=M_Z$, we obtain the value $\alpha_s(M_Z,n_f=5)=0.1159(12)$, which
agrees with the previous result \cite{Maezawa:2016vgv}
but has a larger error. Our result for the central value of $\alpha_s$ is lower than 
many lattice QCD determinations. However, it is only $1.4 \sigma$ 
lower than the PDG value and agrees with the determination
of $\alpha_s$ from the static $Q\bar Q$ energy \cite{Bazavov:2014soa}.

From the sixth, eighth, and tenth moments we determined the charm and bottom quark masses. Our results on the heavy quark masses
agree well with the previous (2+1)-flavor HISQ determination \cite{Maezawa:2016vgv} but have smaller errors.
We also found that our results agree well
with other lattice determinations that are based on various approaches. Thus, our analysis suggests that lattice
determination of the heavy quark masses is under control.

\section*{Acknowledgments}
This work was supported by U.S. Department of Energy under 
Contract No. DE-SC0012704 and
by the DFG cluster of excellence ``Origin and Structure of the Universe'' (www.universe-cluster.de).
The simulations have been carried out on the computing facilities of 
the Computational Center for Particle and Astrophysics (C2PAP) as well as on the SuperMUC at the Leibniz-Rechenzentrum (LRZ).
The lattice QCD calculations have been performed using the publicly available MILC code.
PP would like to thank V. Mateu for useful discussions on the 
the gluon condensate contributions to the moments of quarkonium correlators, K. Nakayama for
correspondence on the lattice results with domain wall fermions and R. Sommer for illuminating
the smallness of the $\alpha_s^3$ coefficient of the reduced moments due to the choice of
the renormalization scale $\mu_m=m_h$. We thank the anonymous referee of this paper for useful suggestions.

\bibliography{pap_bib}

\appendix
\section{Numerical results on the reduced moments}\label{app:A}

In this Appendix we present the 
numerical results on the reduced moments $R_n$.
In Table \ref{tab:allmc} we show 
the numerical results for the reduced moments $R_n$ for $m_h=m_c$. In Tables \ref{tab:R15mc}-\ref{tab:R4mc} 
we present the numerical results for the moments $R_n$ for the larger values of the quark masses, 
$m_h=1.5m_c,~2m_c,~3m_c$, and $4m_c$. In these tables we show three errors for the moments: the statistical errors,
the finite size errors and the errors due to mistuning of the heavy quark mass. The last one was estimated by
fitting the quark mass dependence of the reduced moments by a polynomial and estimating the changes
in the moments from this fit when the heavy quark mass is changed by one sigma. The finite size
errors have been estimated using the free theory calculations as described in the main text.
In Table \ref{tab:allmb}, we give
the results of our calculations of moments at the bottom quark mass.
Finally in Tables \ref{tab:Ratmc}-\ref{tab:Ratmb} we give our numerical results for the ratios $R_6/R_8$ 
and $R_8/R_{10}$ for $m_h=m_c,~1.5m_c,~2m_c,~3m_c,~4m_c$, and $m_b$.
As one can see from Tables \ref{tab:allmc}-\ref{tab:Ratmb} the results for the same $\beta$ but different
light sea quark masses agree within the statistical errors. Thus, the use of heavier than the physical light sea quark masses has
no effect on our results.

\begin{table}
\begin{tabular}{lllllll}
\hline\hline
\(\beta\) & \(\frac{m_\ell}{m_s}\)& \# corr. &  \(R_4\) & \(R_6\) & \(R_8\) & \(R_{10}\) \\
\hline
6.740 & 0.05 & 1601 &  1.19152(15)(11)(30) & 1.02463(8)(27)(11) & 0.94348(5)(8)(20) & 0.89969(4)(16)(23)
\\
6.880 & 0.05 & 1619 &  1.20299(7)(12)(30) & 1.00224(4)(5)(11) & 0.91580(3)(2)(20) & 0.87198(2)(3)(23)
\\
7.030 & 0.05 & 1967 &  1.21414(11)(12)(31) & 0.97833(7)(4)(14) & 0.88936(4)(5)(21) & 0.84685(3)(38)(22)
\\
7.150 & 0.05 & 1317 &  1.22165(14)(13)(37) & 0.96015(7)(5)(18) & 0.87067(4)(3)(25) & 0.82887(4)(0)(26)
\\
7.280 & 0.05 & 1343 &  1.22960(12)(14)(38) & 0.94222(8)(8)(19) & 0.85290(5)(5)(24) & 0.81220(4)(13)(26)
\\
7.373 & 0.05 & 1541 &  1.23459(16)(18)(26) & 0.92688(9)(16)(13) & 0.84089(5)(8)(17) & 0.80120(5)(5)(18)
\\
7.596 & 0.05 & 1585 &  1.24527(18)(9)(10) & 0.90377(9)(43)(5) & 0.81797(6)(213)(6) & 0.78352(4)(678)(6)
\\
7.825 & 0.05 & 1589 &  1.25410(22)(20)(20) & 0.88364(14)(345)(10) & 0.8052(1)(110)(1) & 0.7812(1)(252)(1)
\\
\hline
7.030 & 0.20 & 597 &  1.21440(19)(12)(31) & 0.97833(12)(4)(14) & 0.88924(8)(5)(21) & 0.84666(6)(38)(22) 
\\
7.825 & 0.20 & 298 &  1.25322(39)(20)(20) & 0.88313(21)(345)(10) & 0.8049(1)(110)(1) & 0.7811(1)(252)(1)
\\
8.000 & 0.20 & 462 &  1.26020(69)(112)(50) & 0.8740(3)(109)(2) & 0.8054(2)(271)(2) & 0.7919(2)(509)(1)
\\
8.200 & 0.20 & 487 &  1.27035(65)(398)(28) & 0.8744(4)(297)(1) & 0.8191(3)(581)(0) & 0.8160(2)(916)(0)
\\
8.400 & 0.20 & 495 &  1.27594(115)(945)(350) & 0.8850(6)(596)(1) & 0.8416(4)(980)(5) & 0.845(0)(138)(1)
\\
\hline
\end{tabular}
\caption{Reduced moments for \(m_h=m_c\). 
In the 3rd column we list the number of measurements.
In the 4th through 7th columns we list the moments with the statistical error, 
the finite size error, and the error due to the uncertainty of the quark mass.
}
\label{tab:allmc}
\end{table}
\begin{table}
\begin{tabular}{lllllll}
\hline\hline
\(\beta\) & \(\frac{m_\ell}{m_s}\)& \# corr. & \(R_4\) & \(R_6\) & \(R_8\) & \(R_{10}\) \\
\hline
6.880 & 0.05 & 1619 & 1.13022(4)(12)(35) & 1.03704(2)(5)(17) & 0.98313(1)(3)(38) & 0.95055(1)(2)(48)
\\
7.030 & 0.05 & 1967 & 1.14320(5)(12)(36) & 1.02203(3)(5)(24) & 0.95878(2)(2)(42) & 0.92274(1)(2)(49)
\\
7.150 & 0.05 & 1317 & 1.15241(7)(12)(43) & 1.00802(4)(5)(34) & 0.93945(2)(2)(53) & 0.90251(2)(2)(58)
\\
7.280 & 0.05 & 1343 & 1.16151(9)(12)(43) & 0.99220(6)(5)(37) & 0.91994(4)(2)(54) & 0.88308(3)(2)(57)
\\
7.373 & 0.05 & 723 & 1.16774(12)(12)(30) & 0.98062(8)(5)(27) & 0.90671(5)(2)(37) & 0.87024(4)(1)(39)
\\
7.596 & 0.05 & 642 & 1.18017(16)(9)(11) & 0.95412(9)(47)(18) & 0.87889(5)(12)(14) & 0.84368(4)(25)(14)
\\
7.825 & 0.05 & 627 & 1.19047(15)(14)(24) & 0.93016(10)(1)(22) & 0.85552(6)(42)(25) & 0.82239(6)(177)(21)
\\
\hline
7.030 & 0.20 & 597 & 1.14348(7)(12)(36) & 1.02214(5)(5)(24) & 0.95879(3)(2)(43) & 0.92269(3)(2)(49) 
\\
7.825 & 0.20 & 298 & 1.19049(28)(14)(23) & 0.93003(18)(1)(22) & 0.85539(12)(42)(25) & 0.82228(10)(177)(21)
\\
8.000 & 0.20 & 462 & 1.19752(25)(9)(58) & 0.91320(15)(38)(48) & 0.84039(10)(199)(46) & 0.81108(9)(650)(32)
\\
8.200 & 0.20 & 487 & 1.20592(29)(10)(32) & 0.89663(17)(263)(17) & 0.82892(12)(902)(12) & 0.8079(1)(219)(0)
\\
8.400 & 0.20 & 495 & 1.21132(63)(90)(118) & 0.88617(38)(968)(79) & 0.8279(2)(252)(2) & 0.8171(2)(488)(12)
\\
\hline
\end{tabular}
\caption{Reduced moments for \(m_h=1.5m_c\).
In the 3rd column we list the number of measurements.
In the 4th through 7th columns we list the moments with the statistical error, 
the finite size error, and the error due to the uncertainty of the quark mass.
}
\label{tab:R15mc}
\end{table}
\begin{table}
\begin{tabular}{lllllll}
\hline\hline
\(\beta\) & \(\frac{m_\ell}{m_s}\)& \# corr. & \(R_4\) & \(R_6\) & \(R_8\) & \(R_{10}\) \\
\hline
6.880 & 0.05 & 1619 & 1.108989(3)(11)(24) & 1.04248(2)(5)(2) & 1.01130(1)(3)(16) & 0.99245(1)(2)(26)
\\
7.030 & 0.05 & 1967 & 1.10192(3)(11)(26) & 1.03622(2)(5)(7) & 0.99575(1)(3)(24) & 0.97037(1)(2)(33)
\\
7.150 & 0.05 & 1317 & 1.11140(4)(11)(33) & 1.02855(2)(5)(15) & 0.98066(2)(3)(34) & 0.95131(1)(2)(43)
\\
7.280 & 0.05 & 1343 & 1.12118(6)(12)(34) & 1.01794(4)(5)(20) & 0.96304(3)(2)(38) & 0.93114(2)(2)(45)
\\
7.373 & 0.05 & 723 & 1.12784(9)(12)(24) & 1.00898(6)(5)(16) & 0.95002(4)(2)(28) & 0.91725(3)(2)(31)
\\
7.596 & 0.05 & 625 & 1.14168(8)(11)(9) & 0.98582(5)(5)(8) & 0.92098(3)(2)(11) & 0.88817(2)(1)(12)
\\
7.825 & 0.05 & 627 & 1.15324(13)(11)(19) & 0.96256(9)(5)(17) & 0.89567(6)(2)(22) & 0.86378(4)(3)(23)
\\
\hline
7.030 & 0.20 & 597 & 1.10207(5)(11)(26) & 1.03630(3)(5)(7) & 0.99579(2)(3)(24) & 0.97038(2)(2)(33) 
\\
7.825 & 0.20 & 298 & 1.15332(19)(11)(19) & 0.96265(12)(5)(18) & 0.89573(7)(2)(22) & 0.86383(6)(3)(23)
\\
8.000 & 0.20 & 462 & 1.16101(15)(11)(46) & 0.94509(10)(4)(44) & 0.87797(7)(6)(53) & 0.84709(6)(44)(52)
\\
8.200 & 0.20 & 487 & 1.16967(23)(11)(24) & 0.92603(13)(11)(22) & 0.85960(8)(85)(24) & 0.83121(7)(328)(20)
\\
8.400 & 0.20 & 495 & 1.17615(40)(4)(297) & 0.91003(26)(106)(242) & 0.84664(17)(445)(212) & 0.8239(1)(125)(12)
\\
\hline
\end{tabular}
\caption{Reduced moments for \(m_h=2m_c\).
In the 3rd column we list the number of measurements.
In the 4th through 7th columns we list the moments with the statistical error, 
the finite size error, and the error due to the uncertainty of the quark mass.
}
\label{tab:R2mc}
\end{table}
\begin{table}
\begin{tabular}{lllllll}
\hline\hline
\(\beta\) & \(\frac{m_\ell}{m_s}\)& \# corr. & \(R_4\) & \(R_6\) & \(R_8\) & \(R_{10}\) \\
\hline
6.880 & 0.05 & 1619 & 1.10541(1)(11)(30) & 1.03418(1)(5)(7) & 1.01888(0)(3)(6) & 1.01193(0)(2)(15)
\\
7.030 & 0.05 & 1967 & 1.10597(1)(11)(33) & 1.03505(1)(5)(1) & 1.01754(5)(3)(19) & 1.00841(0)(2)(33)
\\
7.150 & 0.05 & 1317 & 1.06590(3)(11)(27) & 1.03405(2)(5)(1) & 1.01248(1)(3)(18) & 0.99975(1)(2)(29)
\\
7.280 & 0.05 & 1343 & 1.07382(3)(41)(29) & 1.03159(2)(9)(7) & 1.00425(1)(5)(26) & 0.98690(1)(25)(38)
\\
7.373 & 0.05 & 723 & 1.08016(5)(42)(21) & 1.02872(4)(7)(8) & 0.99657(2)(6)(22) & 0.97587(2)(25)(30)
\\
7.596 & 0.05 & 629 & 1.09489(5)(11)(8) & 1.01736(3)(5)(6) & 0.97396(2)(2)(11) & 0.94755(1)(2)(13)
\\
7.825 & 0.05 & 630 & 1.10813(6)(11)(17) & 1.00076(5)(5)(16) & 0.94910(3)(2)(24) & 0.92077(2)(1)(27)
\\
\hline
7.030 & 0.20 & 597 & 1.05968(2)(11)(33) & 1.03507(2)(5)(1) & 1.01755(1)(3)(19) & 1.00841(1)(2)(33) 
\\
7.825 & 0.20 & 298 & 1.10808(10)(11)(17) & 1.00072(7)(5)(16) & 0.94907(4)(2)(24) & 0.92073(4)(1)(27)
\\
8.000 & 0.20 & 462 & 1.11732(10)(84)(41) & 0.98579(7)(51)(45) & 0.93034(5)(2)(61) & 0.90200(4)(37)(65)
\\
8.200 & 0.20 & 487 & 1.12727(14)(90)(22) & 0.96768(10)(63)(26) & 0.90997(6)(14)(32) & 0.88213(5)(48)(33)
\\
8.400 & 0.20 & 495 & 1.13531(24)(50)(114) & 0.95106(16)(27)(324) & 0.89267(10)(3)(375) & 0.86562(9)(18)(351)
\\
\hline
\end{tabular}
\caption{Reduced moments for \(m_h=3m_c\).
In the 3rd column we list the number of measurements.
In the 4th through 7th columns we list the moments with the statistical error, 
the finite size error, and the error due to the uncertainty of the quark mass.
}
\label{tab:R3mc}
\end{table}
\begin{table}
\begin{tabular}{lllllll}
\hline\hline
\(\beta\) & \(\frac{m_\ell}{m_s}\)& \# corr. & \(R_4\) & \(R_6\) & \(R_8\) & \(R_{10}\) \\
\hline
7.150 & 0.05 & 1317 & 1.04559(2)(11)(27) & 1.02832(1)(5)(8) & 1.01534(1)(3)(4) & 1.00922(1)(2)(12)
\\
7.280 & 0.05 & 1343 & 1.05008(2)(41)(22) & 1.02913(1)(9)(4) & 1.01432(1)(5)(6) & 1.00649(1)(25)(14)
\\
7.373 & 0.05 & 818 & 1.05437(3)(42)(17) & 1.02883(2)(7)(1) & 1.01142(1)(6)(8) & 1.00129(1)(25)(15)
\\
7.596 & 0.05 & 629 & 1.06662(3)(10)(7) & 1.02559(2)(5)(2) & 0.99950(1)(2)(6) & 0.98250(1)(2)(9)
\\
7.825 & 0.05 & 630 & 1.07976(5)(10)(15) & 1.01734(3)(5)(8) & 0.98113(2)(2)(19) & 0.95833(2)(2)(23)
\\
\hline
7.825 & 0.20 & 298 & 1.07965(7)(10)(15) & 1.01727(4)(5)(8) & 0.98109(3)(2)(19) & 0.95830(2)(2)(23)
\\
8.000 & 0.20 & 462 & 1.08972(6)(8)(38) & 1.00735(4)(42)(30) & 0.96424(2)(6)(52) & 0.93912(2)(25)(59)
\\
8.200 & 0.20 & 487 & 1.10053(8)(59)(20) & 0.99275(6)(25)(20) & 0.94397(4)(4)(29) & 0.91810(3)(17)(31)
\\
8.400 & 0.20 & 495 & 1.10942(13)(94)(84) & 0.97756(9)(66)(276) & 0.92590(5)(9)(360) & 0.90027(5)(50)(377)
\\
\hline
\end{tabular}
\caption{Reduced moments for \(m_h=4m_c\).
In the 3rd column we list the number of measurements.
In the 4th through 7th columns we list the moments with the statistical error, 
the finite size error, and the error due to the uncertainty of the quark mass.
}
\label{tab:R4mc}
\end{table}
\begin{table}
\begin{tabular}{lllllll}
\hline\hline
\(\beta\) & \(\frac{m_\ell}{m_s}\)& \# corr. &  \(R_4\) & \(R_6\) & \(R_8\) & \(R_{10}\) \\
\hline
7.596 & 0.05 & 638 & 1.05521(2)(66)(6) & 1.02587(1)(39)(0) & 1.00650(1)(2)(3) & 0.99434(1)(3)(6)
\\
7.825 & 0.05 & 641 & 1.06916(3)(26)(6) & 1.02083(2)(1)(2) & 0.99128(1)(7)(5) & 0.97197(1)(14)(7)
\\
\hline
7.825 & 0.20 & 298 & 1.06919(5)(26)(5) & 1.02084(4)(0)(2) & 0.99128(3)(7)(5) & 0.97197(2)(14)(7)
\\
8.000 & 0.20 & 462 & 1.07927(5)(21)(8) & 1.01323(3)(48)(5) & 0.97609(2)(1)(10) & 0.95318(2)(27)(12)
\\
8.200 & 0.20 & 487 & 1.09013(10)(17)(9) & 1.00109(7)(19)(8) & 0.95704(5)(8)(13) & 0.93234(4)(11)(14)
\\
8.400 & 0.20 & 495 & 1.10994(10)(78)(13) & 0.98692(8)(46)(13) & 0.93871(5)(0)(18) & 0.91382(5)(33)(19)
\\
\hline
\end{tabular}
\caption{Reduced moments for \(m_h=m_b\).
In the 3rd column we list the number of measurements.
In the 4th through 7th columns we list the moments with the statistical error, 
the finite size error, and the error due to the uncertainty of the quark mass.
}
\label{tab:allmb}
\end{table}
\begin{table}
\begin{tabular}{lllll}
\hline\hline
\(\beta\) & \(\frac{m_\ell}{m_s}\)& \# corr. &  \(\frac{R_6}{R_8}\) & \(\frac{R_8}{R_{10}}\) \\
\hline
6.740 & 0.05 & 1601 &  1.08601(4)(36)(17) & 1.04867(2)(10)(7) 
\\
6.880 & 0.05 & 1619 &  1.09438(2)(3)(17) & 1.05026(1)(5)(7)
\\
7.030 & 0.05 & 1967 &  1.10004(3)(11)(15) & 1.05020(1)(42)(5)
\\
7.150 & 0.05 & 1317 &  1.10277(4)(3)(16) & 1.05043(1)(4)(5)
\\
7.280 & 0.05 & 1343 &  1.10472(4)(3)(14) & 1.05011(2)(23)(5)
\\
7.373 & 0.05 & 1541 &  1.10560(5)(9)(9) & 1.04955(2)(73)(3)
\\
7.596 & 0.05 & 1585 &  1.10489(5)(235)(3) & 1.04397(2)(637)(0)
\\
7.825 & 0.05 & 1589 &  1.0975(1)(108)(0) & 1.0307(0)(197)(0)
\\
\hline
7.030 & 0.20 & 597 &  1.10018(5)(11)(15) & 1.05029(2)(42)(5)
\\
7.825 & 0.20 & 298 &  1.0972(1)(108)(0) & 1.0305(0)(197)(0)
\\
8.000 & 0.20 & 462 &  1.0851(2)(236)(0) & 1.0170(0)(326)(1)
\\
8.200 & 0.20 & 487 &  1.0676(2)(414)(1) & 1.0038(1)(436)(1)
\\
8.400 & 0.20 & 495 &  1.0515(3)(539)(10) & 0.9964(1)(466)(9)
\\
\hline
\end{tabular}
\caption{Ratios of the reduced moments for \(m_h=m_c\).
In the 3rd column we list the number of measurements.
In the 4th and 5th columns we list the ratios with the statistical error, 
the finite size error, and the error due to the uncertainty of the quark mass.
}
\label{tab:Ratmc}
\end{table}
\begin{table}
\begin{tabular}{lllll}
\hline\hline
\(\beta\) & \(\frac{m_\ell}{m_s}\)& \# corr. & \(\frac{R_6}{R_8}\) & \(\frac{R_8}{R_{10}}\) \\
\hline
6.880 & 0.05 & 1619 & 1.05484(1)(3)(20) & 1.03428(0)(1)(10)
\\
7.030 & 0.05 & 1967 & 1.06596(1)(3)(19) & 1.03906(1)(1)(8)
\\
7.150 & 0.05 & 1317 & 1.07299(2)(3)(21) & 1.04094(1)(1)(8)
\\
7.280 & 0.05 & 1343 & 1.07855(2)(3)(19) & 1.04173(1)(1)(6)
\\
7.373 & 0.05 & 723 & 1.08151(3)(3)(12) & 1.04192(2)(1)(4)
\\
7.596 & 0.05 & 642 & 1.08560(5)(75)(4) & 1.04173(2)(16)(1)
\\
7.825 & 0.05 & 627 & 1.08724(4)(57)(5) & 1.04028(2)(176)(2)
\\
\hline
7.030 & 0.20 & 597 & 1.06607(2)(3)(19) & 1.03913(1)(1)(8)
\\
7.825 & 0.20 & 298 & 1.08726(7)(57)(5) & 1.04026(3)(176)(2)
\\
8.000 & 0.20 & 462 & 1.08663(7)(212)(5) & 1.03614(3)(590)(12)
\\
8.200 & 0.20 & 487 & 1.08169(8)(868)(3) & 1.0261(0)(170)(1)
\\
8.400 & 0.20 & 495 & 1.0703(2)(213)(10) & 1.0132(1)(308)(8)
\\
\hline
\end{tabular}
\caption{Ratios of the reduced moments for \(m_h=1.5m_c\).
In the 3rd column we list the number of measurements.
In the 4th and 5th columns we list the ratios with the statistical error, 
the finite size error, and the error due to the uncertainty of the quark mass.
}
\label{tab:Rat15mc}
\end{table}
\begin{table}
\begin{tabular}{lllll}
\hline\hline
\(\beta\) & \(\frac{m_\ell}{m_s}\)& \# corr. & \(\frac{R_6}{R_8}\) & \(\frac{R_8}{R_{10}}\) \\
\hline
6.880 & 0.05 & 1619 & 1.03082(1)(3)(13) & 1.01900(0)(1)(9)
\\
7.030 & 0.05 & 1967 & 1.04064(1)(3)(16) & 1.02616(0)(1)(9)
\\
7.150 & 0.05 & 1317 & 1.04884(1)(3)(20) & 1.03085(0)(1)(10)
\\
7.280 & 0.05 & 1343 & 1.05700(2)(3)(19) & 1.03426(1)(1)(9)
\\
7.373 & 0.05 & 723 & 1.06207(2)(3)(13) & 1.03573(1)(1)(5)
\\
7.596 & 0.05 & 625 & 1.07040(2)(3)(4) & 1.03695(1)(1)(1)
\\
7.825 & 0.05 & 627 & 1.07469(3)(3)(7) & 1.03692(2)(6)(2)
\\
\hline
7.030 & 0.20 & 597 & 1.04068(1)(3)(16) & 1.02618(1)(1)(9)
\\
7.825 & 0.20 & 298 & 1.07471(6)(3)(7) & 1.03692(3)(6)(2)
\\
8.000 & 0.20 & 462 & 1.07644(4)(12)(15) & 1.03646(2)(46)(2)
\\
8.200 & 0.20 & 487 & 1.07729(7)(94)(5) & 1.03415(3)(307)(3)
\\
8.400 & 0.20 & 495 & 1.07488(11)(441)(1) & 1.0276(0)(104)(9)
\\
\hline
\end{tabular}
\caption{Ratios of the reduced moments for \(m_h=2m_c\).
In the 3rd column we list the number of measurements.
In the 4th and 5th columns we list the ratios with the statistical error, 
the finite size error, and the error due to the uncertainty of the quark mass.
}
\label{tab:Rat2mc}
\end{table}
\begin{table}
\begin{tabular}{lllll}
\hline\hline
\(\beta\) & \(\frac{m_\ell}{m_s}\)& \# corr. & \(\frac{R_6}{R_8}\) & \(\frac{R_8}{R_{10}}\) \\
\hline
6.880 & 0.05 & 1619 & 1.01502(0)(3)(11) & 1.00687(0)(1)(8)
\\
7.030 & 0.05 & 1967 & 1.01721(0)(3)(16) & 1.00905(0)(1)(11)
\\
7.150 & 0.05 & 1317 & 1.02131(1)(3)(14) & 1.01273(0)(1)(10)
\\
7.280 & 0.05 & 1343 & 1.02722(1)(6)(17) & 1.01759(0)(23)(11)
\\
7.373 & 0.05 & 723 & 1.03226(1)(13)(12) & 1.02122(0)(21)(7)
\\
7.596 & 0.05 & 629 & 1.04456(1)(2)(5) & 1.02787(0)(1)(2)
\\
7.825 & 0.05 & 630 & 1.05444(2)(3)(9) & 1.03077(1)(1)(3)
\\
\hline
7.030 & 0.20 & 597 & 1.01722(1)(3)(16) & 1.00906(0)(1)(11)
\\
7.825 & 0.20 & 298 & 1.05443(3)(3)(9) & 1.0309(0)(7)
\\
8.000 & 0.20 & 462 & 1.05960(3)(70)(13) & 1.03141(1)(42)(6)
\\
8.200 & 0.20 & 487 & 1.06343(4)(97)(8) & 1.03156(2)(41)(2)
\\
8.400 & 0.20 & 495 & 1.06541(8)(37)(13) & 1.03125(3)(26)(4)
\\
\hline
\end{tabular}
\caption{Ratios of the reduced moments for \(m_h=3m_c\).
In the 3rd column we list the number of measurements.
In the 4th and 5th columns we list the ratios with the statistical error, 
the finite size error, and the error due to the uncertainty of the quark mass.
}
\label{tab:Rat3mc}
\end{table}
\begin{table}
\begin{tabular}{lllll}
\hline\hline
\(\beta\) & \(\frac{m_\ell}{m_s}\)& \# corr. & \(\frac{R_6}{R_8}\) & \(\frac{R_8}{R_{10}}\) \\
\hline
7.150 & 0.05 & 1317 & 1.01279(0)(3)(10) & 1.00606(0)(1)(7)
\\
7.280 & 0.05 & 1343 & 1.01460(0)(6)(10) & 1.00778(0)(23)(7)
\\
7.373 & 0.05 & 818 & 1.001721(1)(7)(9) & 1.01012(0)(6)(6)
\\
7.596 & 0.05 & 629 & 1.02610(1)(2)(4) & 1.01731(0)(1)(3)
\\
7.825 & 0.05 & 630 & 1.03691(1)(3)(10) & 1.02379(0)(1)(5)
\\
\hline
7.825 & 0.20 & 298 & 1.03688(2)(3)(10) & 1.02378(1)(1)(5)
\\
8.000 & 0.20 & 462 & 1.04472(2)(53)(22) & 1.02675(1)(23)(8)
\\
8.200 & 0.20 & 487 & 1.05167(2)(31)(10) & 1.02818(1)(26)(3)
\\
8.400 & 0.20 & 495 & 1.05580(3)(95)(99) & 1.02847(2)(49)(28)
\\
\hline
\end{tabular}
\caption{Ratios for \(m_h=4m_c\).
In the 3rd column we list the number of measurements.
In the 4th and 5th columns we list the ratios with the statistical error, the finite size error,
and the error due to the uncertainty of the quark mass.
}
\label{tab:Rat4mc}
\end{table}
\begin{table}
\begin{tabular}{lllll}
\hline\hline
\(\beta\) & \(\frac{m_\ell}{m_s}\)& \# corr. &  \(\frac{R_6}{R_8}\) & \(\frac{R_8}{R_{10}}\) \\
\hline
7.596 & 0.05 & 638 & 1.01924(1)(47)(3) & 1.01223(0)(2)(3)
\\
7.825 & 0.05 & 641 & 1.02981(1)(5)(4) & 1.01986(3)(7)(2)
\\
\hline
7.825 & 0.20 & 298 & 1.02982(1)(5)(4) & 1.01987(0)(7)(2)
\\
8.000 & 0.20 & 462 & 1.03805(1)(55)(5) & 1.02404(1)(33)(2)
\\
8.200 & 0.20 & 487 & 1.04602(3)(26)(5) & 1.02649(1)(3)(2)
\\
8.400 & 0.20 & 495 & 1.05136(3)(62)(6) & 1.02724(1)(39)(2)
\\
\hline
\end{tabular}
\caption{Ratios of the reduced moments for \(m_h=m_b\).
In the 3rd column we list the number of measurements.
In the 4th and 5th columns we list the ratios with the statistical error, 
the finite size error, and the error due to the uncertainty of the quark mass.
}
\label{tab:Ratmb}
\end{table}

\section{Details of continuum extrapolations}\label{app:B}

In this Appendix, we discuss further details of the continuum extrapolations.
As mentioned in the main text, we performed a variety of continuum extrapolations
using Eq. \eqref{fit} and keeping a different number of terms in the sum. In doing so,
we varied the fit interval such that the $\chi^2/df$ was close to or below one. Fewer terms in
Eq. (\ref{fit}) usually means a more restricted interval in $a m_{h0}$. 

We first discuss the continuum extrapolation of the fourth reduced moment, $R_4$,
which is the most challenging. Here we find that including terms up to $N=2$ in Eq. (\ref{fit})
is important if we want to obtain good fits in extended region in $a m_{h0}$. For $m_h \le 2 m_c$,
the coefficient $c_{21}$ in Eq. (\ref{fit}) could be treated as free fit parameter 
as we have many data points for relatively small $a m_{h0}$ to have stable fits.
We considered constrained fits, in which $c_{2j}/c_{1j}$ was fixed to some value between $-4$ and $-5.5$ but
the continuum extrapolated $R_4$ value did not change much. 
To study the effect of higher order terms in $\alpha_s$ on the continuum
extrapolations we also included terms proportional to $\alpha_s^3$ in the fit with coefficient
$c_{3j}=c_{1j}$. The finite volume errors are sizable 
for the three finest lattices, see e.g. Fig. \ref{fig:r4_rat8}.
To exclude the possibility
that underestimated finite volume errors affect the continuum extrapolation
we performed fits omitting the data points corresponding to the two smallest lattice spacings,
where finite volume effects are the largest. We find that doing so does
not affect the final continuum estimate within errors.
For $m_h=3m_c$ we do not have enough data points to treat $c_{21}$ as free fit parameters, so
we performed constrained fits with $c_{2j}=-5 c_{2j}$ and standard fits with
using $c_{2j}=0$. The two types of fits gave consistent results, see below.

In Fig. \ref{fig:contR4} we show continuum estimates for $R_4$ obtained from
different fits that are labeled as $NJ_{max2}$, with $N$ and $J$ being the number
of terms in Eq. (\ref{fit}) and $max2$ being the maximal value of $(a m_{h0})^2$ that
enters the fit, i.e., $(a m_{h0})^2 \le max2$. Constrained fits are indicated by a subscript $c$.
Furthermore, the additional restrictions on the 
beta values used in the fits are also marked in the legend.
The central values of the continuum estimates show some scattering, although most
of them agree within errors.
We take the weighted average of these estimates to obtain our
final continuum result for $R_4$, which is shown as the solid horizontal line 
in Fig. \ref{fig:contR4}. We then assign an error to this continuum result.
The error band was determined in the following way. When the scattering of individual fits
was comparable to the corresponding errors the error band was defined such that all individual
fits agreed with the final continuum estimate within errors. When the errors of different fits
are much larger than the scattering of the corresponding central values the error band represents
the typical error of the fits.
The errors on the final continuum estimate are
represented by dashed vertical bands in Fig. \ref{fig:contR4}.
We estimated the finite volume errors on
the reduced moments using the free theory result, see above. 
As one can see from Fig. \ref{fig:r4_rat8} the finite size effects estimated
this way are fairly large for the three finest lattices.
To exclude the possibility
that finite volume errors are underestimated
we also performed fits omitting the data points corresponding to the three smallest lattice spacings
for $m_h=m_c$, where finite volume effects are the largest. We find that doing so does
not affect the final continuum estimate within errors. 
\begin{figure}
\includegraphics[width=4cm]{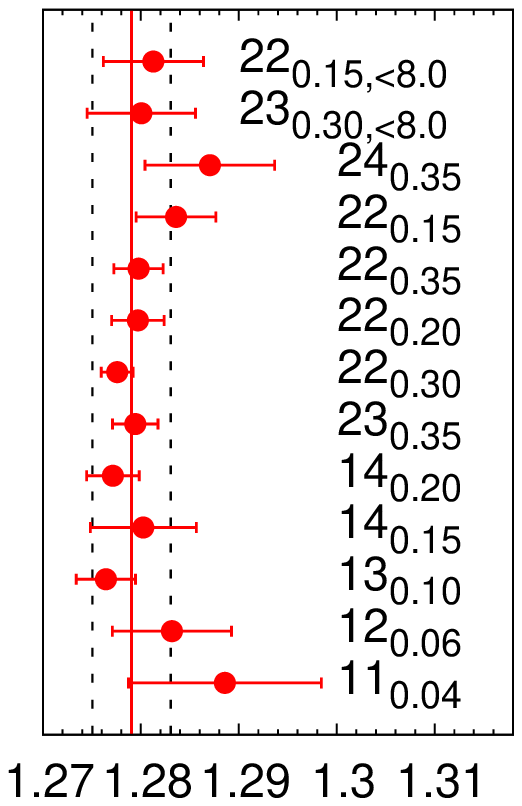}
\includegraphics[width=4cm]{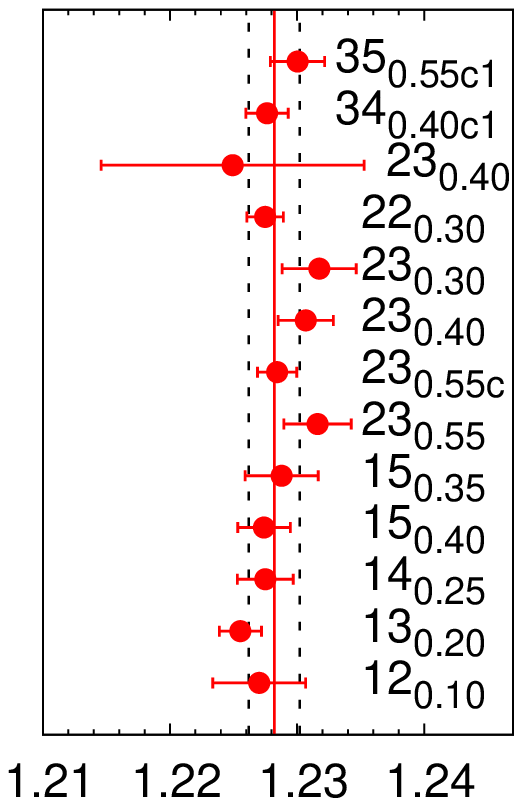}
\includegraphics[width=4cm]{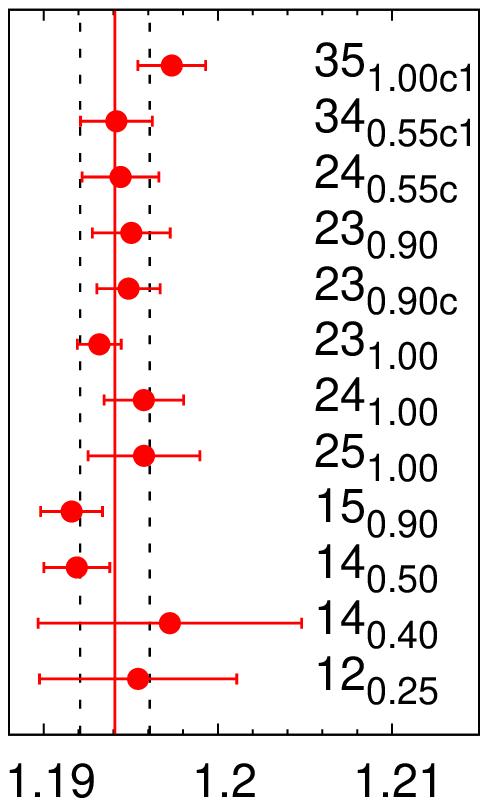}
\includegraphics[width=4cm]{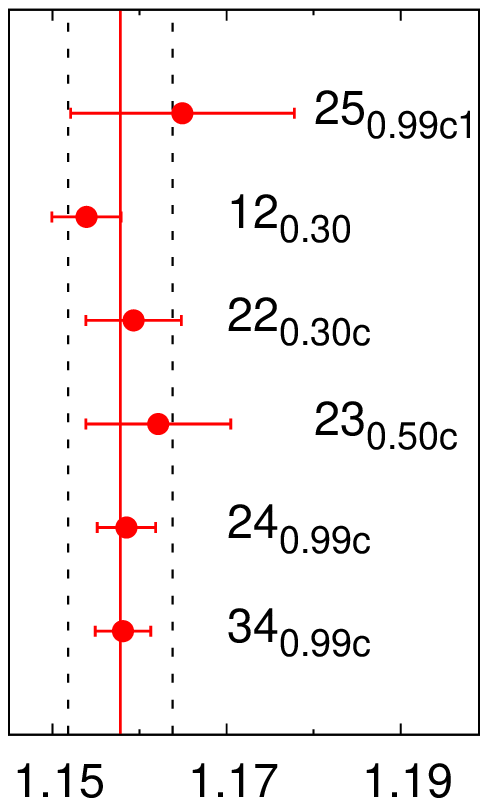}
\caption{The comparison of different continuum extrapolations for $R_4$. The panels from left to right
show the results for $m_h=m_c,~1.5m_c,~2m_c$, and $3m_c$. The solid vertical line corresponds
to the weighted average of different fits, while the vertical dashed lines indicate the estimated
uncertainty of the continuum value of $R_4$. The subscripts $c$ and $c1$ indicate the constrained fits
with $c_{2j}=-5 c_{1j}$ and with $c_{2j}=-5 c_{1j},~c_{3j}=c_{1j}$, respectively.
In the leftmost panel we also show fits that only use data with $\beta<8.0$.
Individual continuum fits are labeled as $NJ_{max2}$ with $N$ and $J$ denoting the number
of terms in Eq. (\ref{fit}) and $max2$ being the maximal value of $(a m_{h0})^2$, see text.
}
\label{fig:contR4}
\end{figure}

\begin{figure}
\includegraphics[width=4cm]{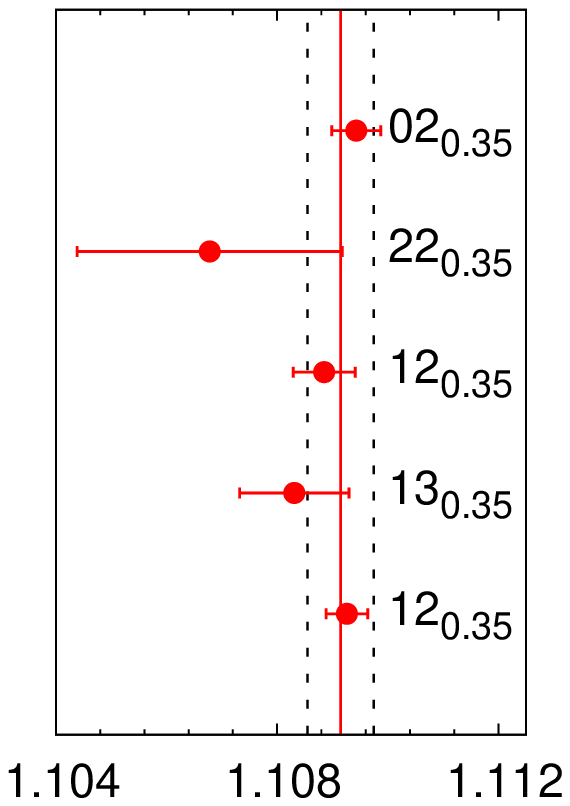}
\includegraphics[width=4cm]{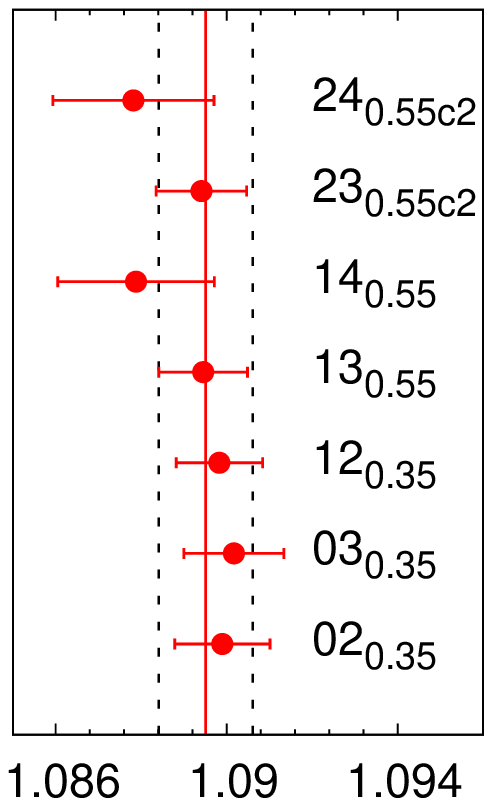}
\includegraphics[width=4cm]{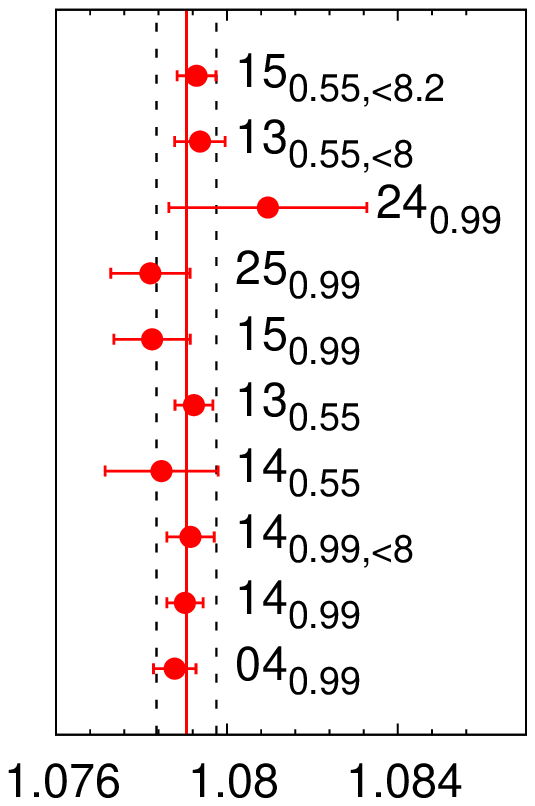}
\includegraphics[width=4cm]{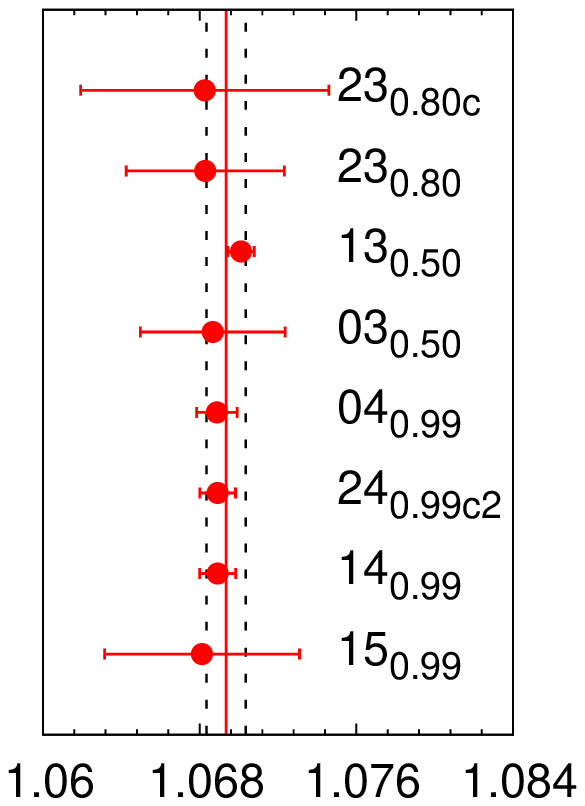}
\caption{The comparison of different continuum extrapolations for $R_6/R_8$. The panels from left to right
show the results for $m_h=m_c,~1.5m_c,~2m_c$, and $3m_c$. The solid vertical line corresponds
to the weighted average of different fits, while the vertical dashed lines indicate the estimated
uncertainty of the continuum value of $R_6/R_8$.
Subscripts $c$ and $c2$ correspond to the constrained fits with $c_{2j}=-5 c_{1j}$ and with
$c_{2j}=c_{1j}$, respectively. In the third panel we also show fits that only use data with $\beta<8.0$
and $\beta<8.2$. We use the same labeling scheme of  individual continuum fits as in Fig. \ref{fig:contR4}.
}
\label{fig:contRat6}
\end{figure}

\begin{figure}
\includegraphics[width=4cm]{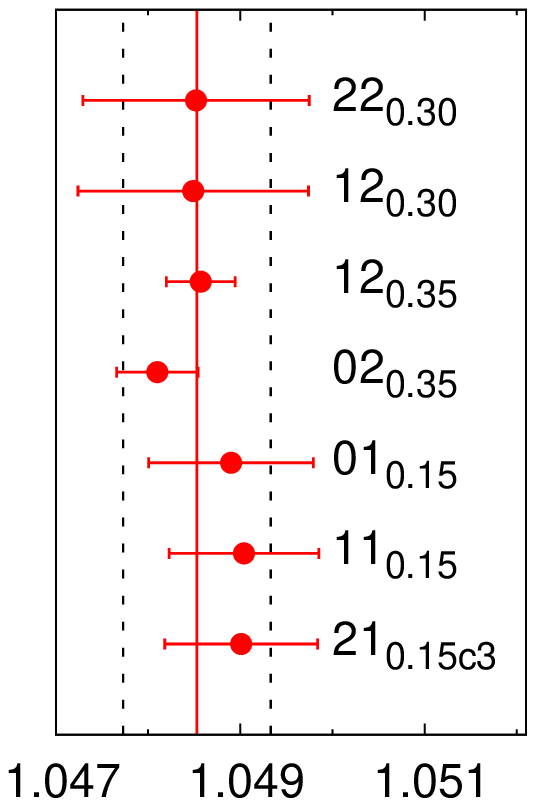}
\includegraphics[width=4cm]{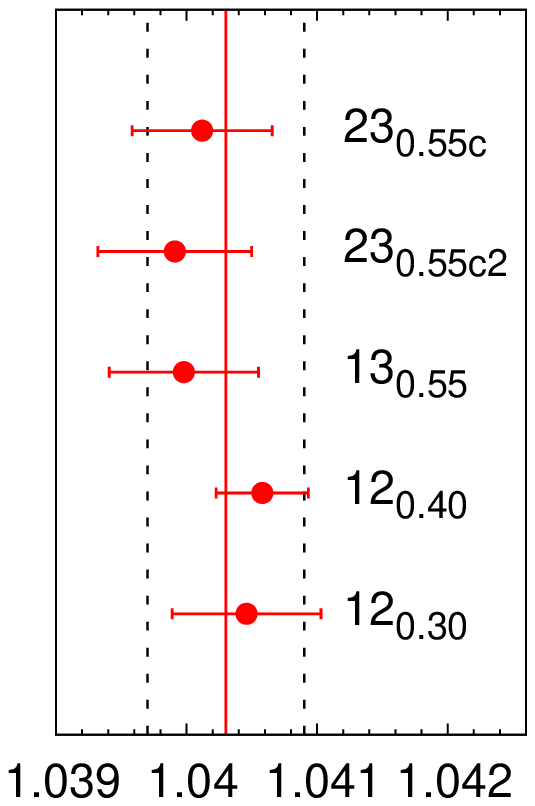}
\includegraphics[width=4cm]{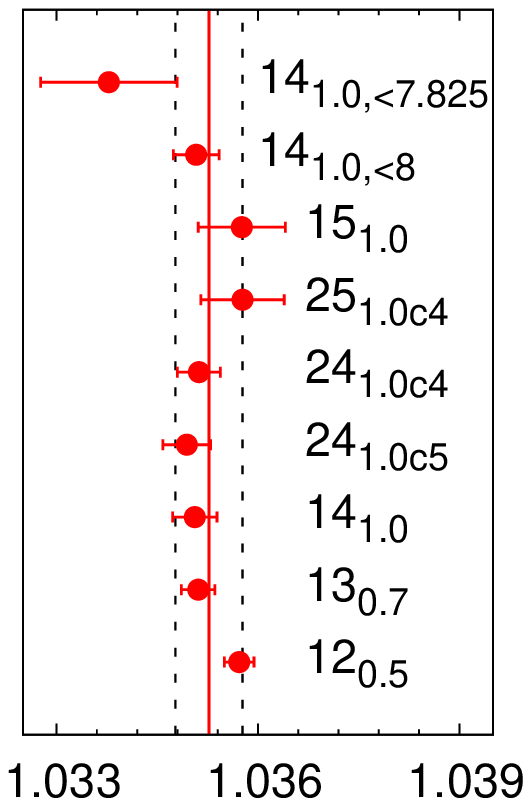}
\includegraphics[width=4cm]{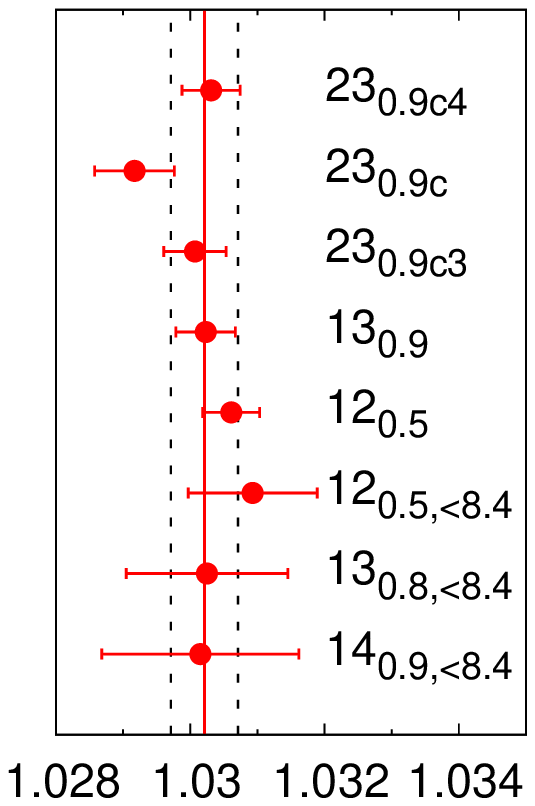}
\caption{The comparison of different continuum extrapolations for $R_8/R_{10}$. The panels from left to right
show the results for $m_h=m_c,~1.5m_c,~2m_c$, and $3m_c$. The solid vertical line corresponds
to the weighted average of different fits, while the vertical dashed lines indicate the estimated
uncertainty of the continuum value of $R_8/R_{10}$.
Subscripts $c$, $c2$, $c_3$, $c4$, and $c5$ correspond to the constrained fits with $c_{2j}=-5 c_{1j}$, with
$c_{2j}=c_{1j}$, with $c_{2j}=-c_{1j}$, with $c_{2j}=2 c_{1j}$, and with $c_{2j}=-2 c_{1j}$, respectively.
We use the same labeling scheme of  individual continuum fits as in Fig. \ref{fig:contR4}.
}
\label{fig:contRat8}
\end{figure}

A similar analysis has been performed also for the ratios of the reduced moments, $R_6/R_8$, and $R_8/R_{10}$.
Here including terms up to $N=2$ in Eq. (\ref{fit}) turned out to be less
important and good fits could be obtained already with $N=1$ in the entire range of lattice spacings.
The corresponding results are shown in Fig. \ref{fig:contRat6} and Fig. \ref{fig:contRat8}.
We use the same labeling scheme for different fits as in the Fig. \ref{fig:contR4}. Here
we also consider constrained fits with $c_{2j}=c_{1j}$, $c_{2j}=-c_{1j}$, $c_{2j}=2 c_{1j}$,
and $c_{2j}=-2 c_{1j}$ labeled as $c2$, $c3$, $c4$, and $c5$, respectively.
To rule out the possibility that the continuum extrapolations are affected by underestimated
finite volume errors we perform the fits omitting data points at large $\beta$ (corresponding to fine lattices).

\begin{figure}
\includegraphics[width=4cm]{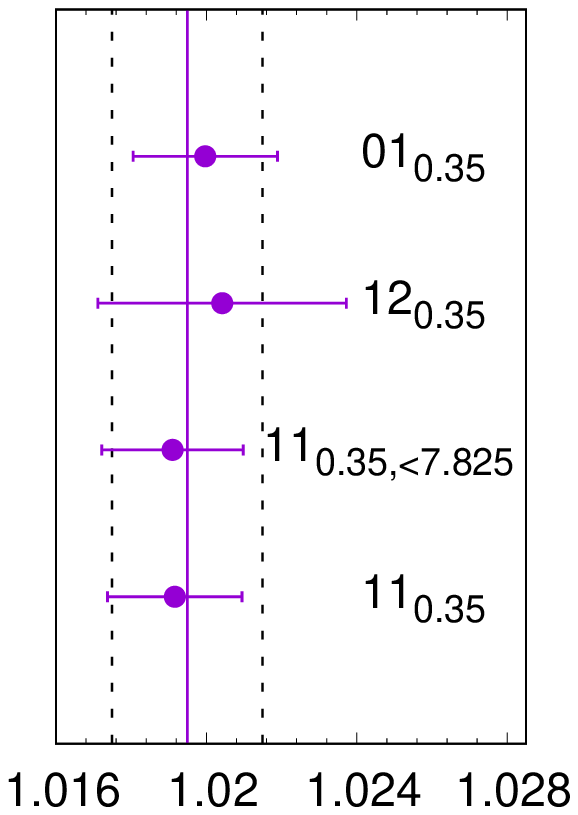}
\includegraphics[width=4cm]{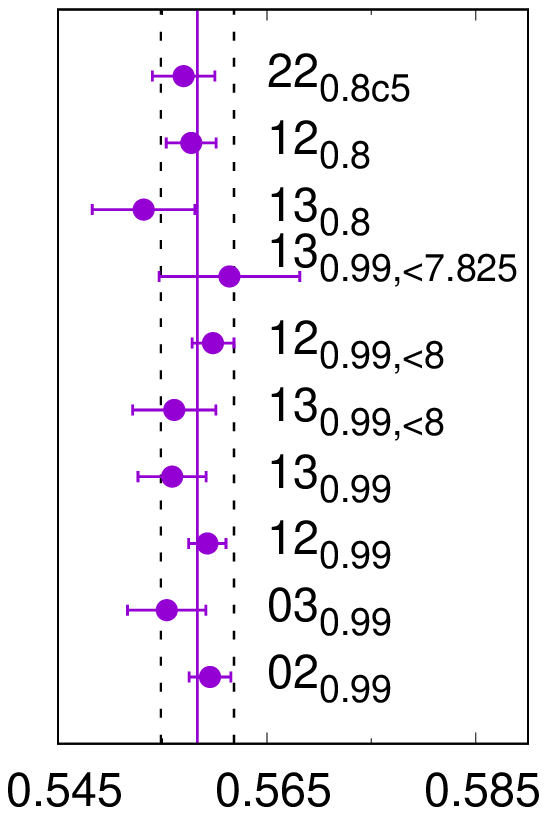}
\includegraphics[width=4cm]{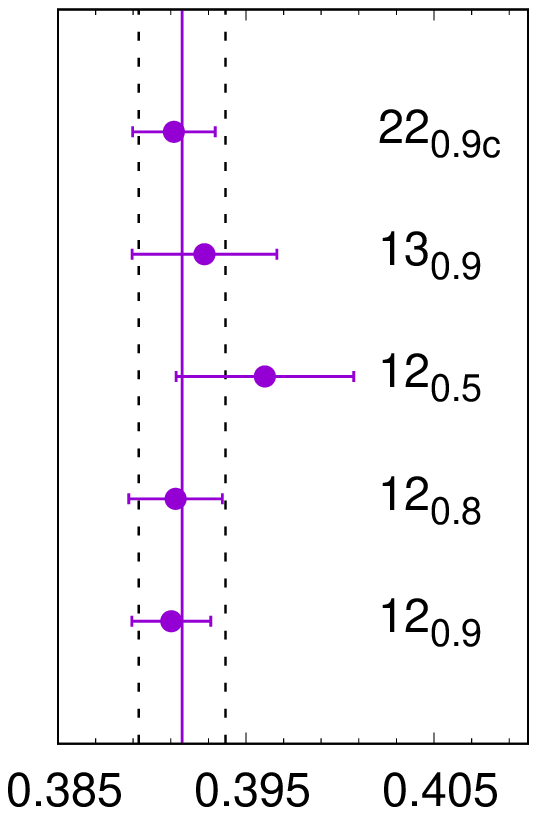}
\includegraphics[width=4cm]{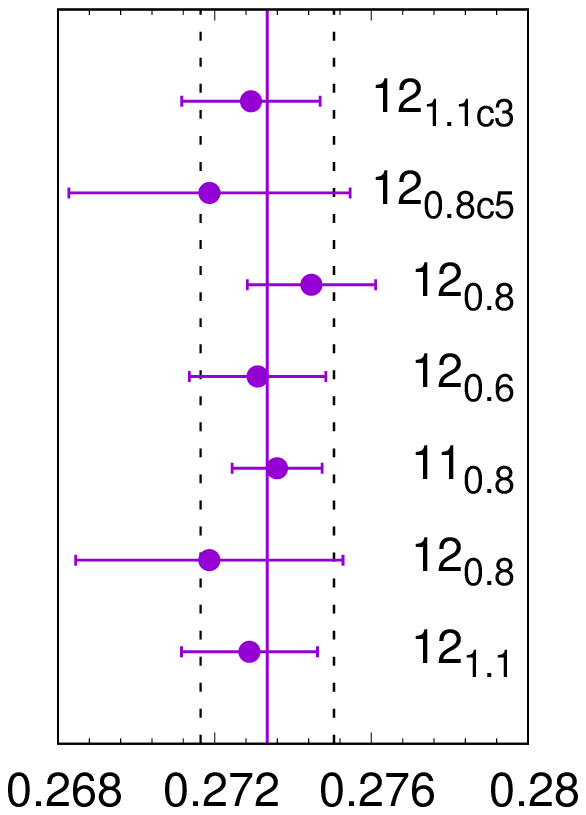}
\caption{The comparison of different continuum extrapolations for $R_6/m_{h0}$. The panels from left to right
show the results for $m_h=m_c,~2m_c,~3m_c$, and $m_b$. The solid vertical line corresponds
to the weighted average of different fits, while the vertical dashed lines indicate the estimated
uncertainty of the continuum value of $R_6/m_{h0}$.
We use the same labeling scheme of  individual continuum fits as in Fig. \ref{fig:contR4}.
}
\label{fig:contR6}
\end{figure}

In Fig. \ref{fig:contR6} we show different continuum extrapolations for $R_6/m_{h0}$
at $m_h=m_c,~2m_c,~3 m_c$, and $m_b$. The results for $m_h=1.5m_c$ and $4m_c$ are not shown,
as they look very similar. 
Here, including terms with $N=2$ in the fit is not important and in many cases
simple, $\alpha_s (a^2+a^4)$ fits do an excellent job. For $m_h=m_c$ even the simplest
$a^2$ fit works very well. Nevertheless, to check for possible systematic
effects we also performed constrained fits that use more terms in Eq. (\ref{fit}).
More specifically, we used constrained fits $c$, $c3$, and $c5$ described above.
For the four smallest lattice spacing the finite volume errors are very large for $R_n,~n\ge 6$, as one can see
from Tables \ref{tab:allmc}-\ref{tab:R4mc}. Because of this the corresponding data do not influence the fit much.
The FLAG review requires that the physical box size $L$ should satisfy $m_{\pi} L>3$ for the finite
volume effects to be acceptable. If we use only the lattice spacings that satisfy this criterion
we obtain $R_6/m_{c0}=1.018(55)$, which agrees with the continuum value that includes all
the data points. This is due to the fact that the cutoff dependence of $R_6/m_{c0}$
is well described by the $a^2$ form.

To rule out the possibility that underestimated finite volume errors influence the continuum result, we also
carried out fits using only data with $\beta<8.0$ and $\beta<7.825$, which are shown in Fig. \ref{fig:contR6}.
We performed the weighted average of different continuum extrapolation
to obtain our final continuum result for $R_6/m_{h0}$, shown as the vertical line in the figure.
The error band of this final continuum result is indicated by the dashed lines in Fig. \ref{fig:contR6}.
Since the continuum estimates from different fits do not scatter much, the error band is mostly
given by the typical error of the individual fits. 
The same analysis was performed also for $R_8/m_{h0}$ and $R_{10}/m_{h0}$ and the results look
very similar. Therefore, we do not shown them here.

\end{document}